\documentclass[journal]{IEEEtran}
\usepackage{graphicx}
\usepackage{epstopdf}
\usepackage{multirow}
\usepackage{booktabs}
\usepackage{multicol, blindtext}
\usepackage{tabularx}
\usepackage{cite}
\usepackage{caption}
\usepackage{gensymb}
\usepackage{amsmath}
\usepackage{array}
\usepackage{float}
\usepackage{xcolor}
\usepackage{color, colortbl}

\usepackage[nottoc]{tocbibind}
 
\ifCLASSOPTIONcompsoc
\usepackage[caption=false,font=normalsize,labelfont=sf,textfont=sf]{subfig}
\else
  \usepackage[caption=false,font=footnotesize]{subfig}
\fi

\usepackage{dblfloatfix}
\usepackage[]{authblk}
\usepackage{url}

\author[1]{Sagar Dutta, \IEEEmembership{Graduate Student Member,~IEEE,}}
\author[1]{Banani Basu, \IEEEmembership{Senior Member,~IEEE,}}
\author[1]{Fazal Ahmed Talukdar, \IEEEmembership{Senior Member,~IEEE}}

\affil[1]{Department of Electronics and Communication Engineering \protect\\ National Institute Of Technology, Silchar, Assam, India}

\begin{document}

\title{Classification of Induction Motor Fault and Imbalance based on Vibration Signal using Single Antenna's Reactive Near-Field}
\maketitle

\begin{abstract}
Early fault diagnosis is imperative for the proper functioning of rotating machines. It can reduce economic losses in the industry due to unexpected failures. Existing fault analysis methods are either expensive or demand expertise for the installation of the sensors. This paper proposes a novel method for the detection of bearing faults and imbalance in induction motors using an antenna as the sensor which is non-invasive and cost-efficient. Time-varying S11 is measured using an omnidirectional antenna and it is seen that the spectrogram of S11 shows unique characteristics for different fault conditions. The experimental setup has analytically evaluated the vibration frequencies due to fault and validated the characteristic fault frequency by applying FFT analysis on the captured $S_{11}$ data. The paper has evaluated the average power content of the detected signals at normal and different fault conditions. A deep learning model is used to classify the faults based on the reflection coefficient ($S_{11}$). It is found that classification accuracy of 98.2\% is achieved using both magnitude and phase of S11, 96\% using the magnitude of $S_{11}$ and 92.1\% using the phase of $S_{11}$. The classification accuracy for different operating frequencies, antenna location and time window are also investigated.
\end{abstract}


\section{introduction}

The induction motor due to its robustness, high reliability and cost-effectiveness is largely used in the industry. Different industrial applications such as electric cars, machine tools and pumps use induction motors. Therefore, it is very important for the industry to accurately track induction motor's health to recognize the early signs of any defect and to prevent unforeseen failures that could lead to significant productivity and economic losses. Numerous methods have been developed for the detection of the fault in the electrical machinery. Bearing faults are the most frequent faults in an electric motor of all mechanical failures, followed by stator faults and rotor faults \cite{rotor1,rotor2,rotor3}.

With the ongoing fourth industrial revolution, new smart technology is being used to automate traditional manufacturing and industrial practices. It includes cyber-physical systems (CPS) \cite{cyber} which is an internet-enabled entity embedded with computing and communicating infrastructures consisting of sensors and actuators. In the years to come, the industries will begin to leverage CPS to achieve operational excellence. This would demand a large number of sensors and the cost of these sensors would account for a significant portion of the overall system cost. Fortunately, advances in manufacturing technology are helping to lower sensor costs. However, most traditional sensors are invasive and it is often complicated to install and maintain. In comparison, the non-invasive sensor requires no expertise or professional tool for the installation process and does not disrupt the equipment's operation, which is cost-effective. The above advantages makes non-invasive measurement  an attractive research direction \cite{research1,research2,research3}.

The vibration of mechanical equipment can be measured by many kinds of sensors, each with its own set of benefits and drawbacks when used in diverse applications. The most widely used vibration sensor is the piezoelectric accelerometer. It's simple to set up and, due to its small size, it's suitable for a wide range of applications. For example, Medeiros et al. successfully measured the flow rate of water by measuring the vibration of the pipes using a piezoelectric accelerometer without intrusion \cite{pipe}. However, in some cases, when the mass of the measured object is very small, the attachment of the sensor may affect the overall characteristic of the original vibration. Furthermore, utilizing current frequency spectral subtraction, motor parameter such as stator current is used to locate bearing defects in induction motor \cite{mcsa}. Esfahani et al. proposed a wireless sensor networks that uses vibration, current, and acoustic sensors to continuously monitor bearing faults\cite{vca}. Rotor speed-based bearing fault diagnosis is another contact-based method that utilizes absolute value-based principal component analysis technique \cite{rotor}. These approaches disrupt machine performance while setting up the sensors, and are impacted by machine temperature and other harsh environmental changes.

In such situations, non-contact monitoring approaches are more appropriate. Zhou et al. \cite{optical1} proposed a laser vibration measurement system that is based on the optical Doppler shift principle to measure regular and irregular vibrating targets. A motion-induced eddy current sensor is proposed by Xue et al. \cite{eddy} that can measure the vibration of non-ferromagnetic/ferromagnetic metals. Other magnetic fields or surrounding ferromagnetic materials, however, can have an effect on it. Sound analysis \cite{sound1} is another method that uses a microphone to record the sound caused by vibrations. However, the background noise due to other machines can distort the bearing noise if not shielded. Therefore, this method is not useful where there are other machines nearby. A high-frequency ultra wide-band radar is proposed by Barusu et. al \cite{radar} to identify faults in the motor using the reflected signal from the motor. The transmitted signal from the radar may interact with a variety of targets causing undesirable interfering noise. Frosini et al. utilized a flux sensor coil to measure the stray flux around the induction motor to identify faults \cite{flux}. The sensor coil can pick up the flux of the nearby electrical machines if they are placed close to each other. Despite the fact that these sensors can sense vibrations with great precision and without contact, but due to their high cost and strict working environments their application is limited. A non-intrusive vibration sensor should be cost-effective, simple to install and suitable for tough environments.

\begin{figure}[!t]
\centering
\subfloat[]{\includegraphics[width=2.2in]{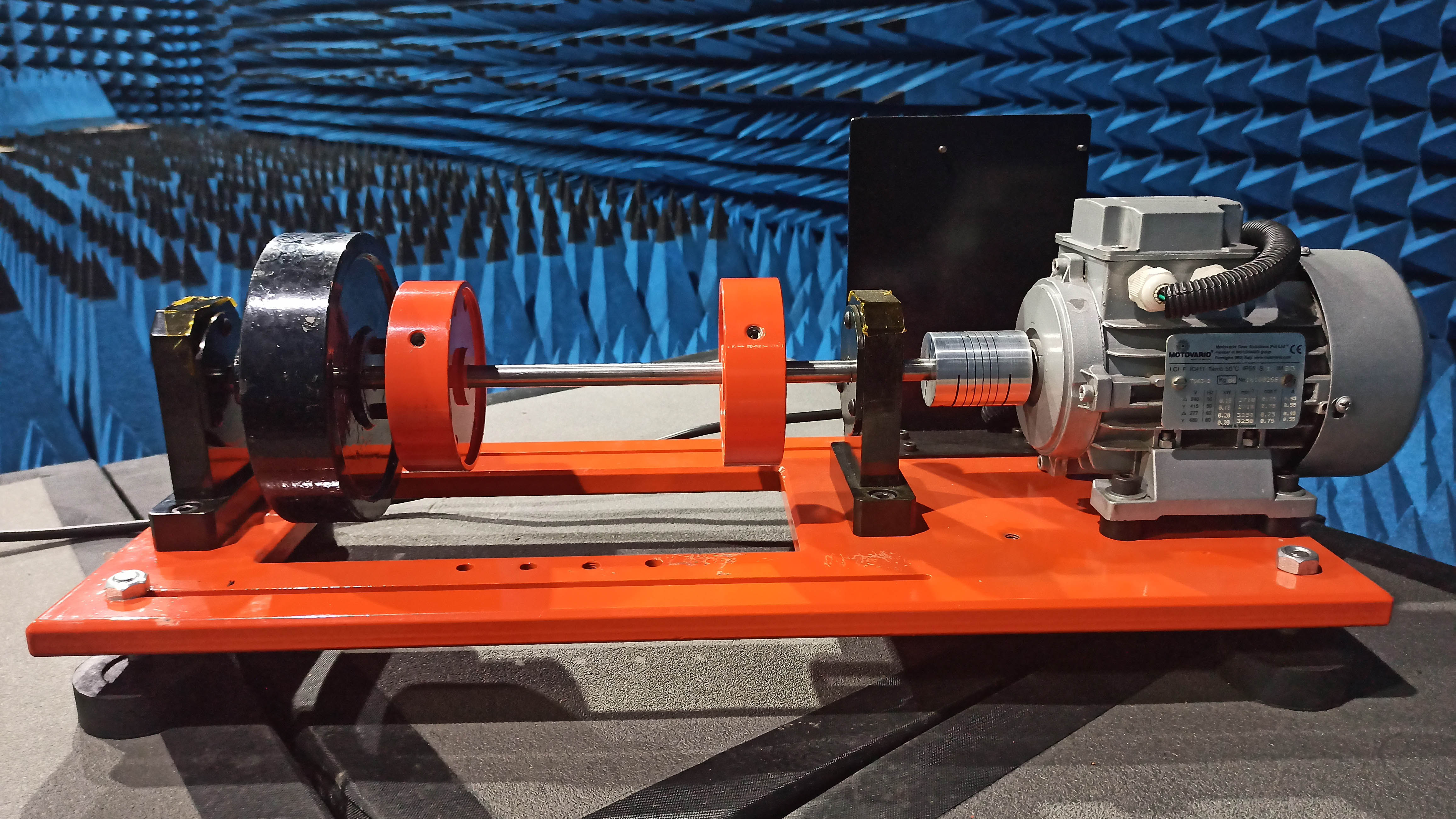}%
\label{setup_1a}}

\subfloat[]{\includegraphics[width=2.2in]{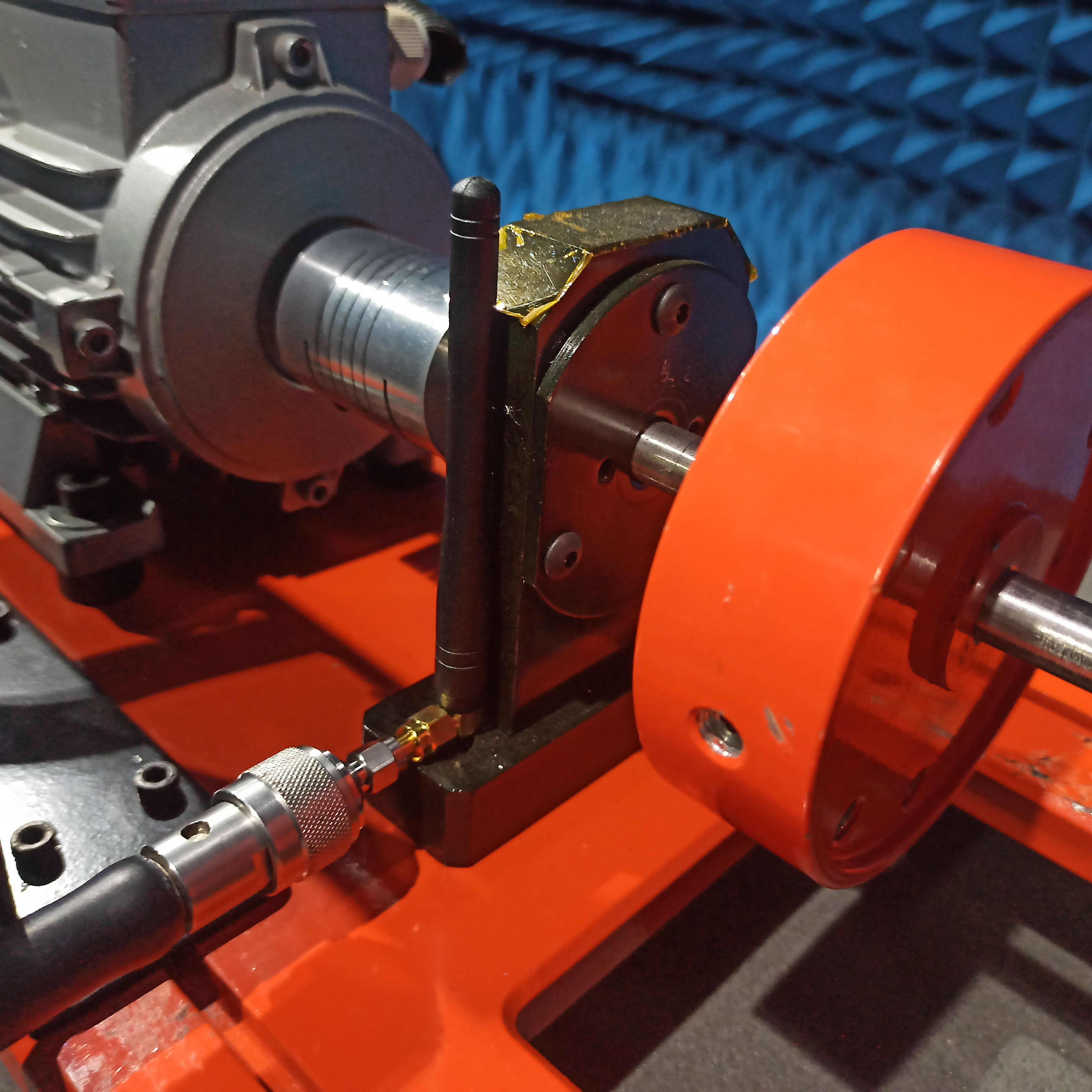}%
\label{setup_1b}}
\caption{(a) Experimental Setup (b) Placement of antenna for reflection coefficient measurement}
\label{setup}
\end{figure}

Moreover, it is studied that the deep learning approach has a significant impact on image processing and pattern recognition in recent years. To diagnose motor faults, deep learning approaches such as neural-fuzzy logic \cite{fuzzy}, deep auto-encoder network \cite{auto}, artificial neural network (ANN) \cite{ann} and convolutional neural network (CNN) \cite{cnn} have been reported. The deep learning approach combined with the microwave systems has gained a substantial attraction of the researchers due to its simplicity and robustness in classifying the microwave signal and applied in the area of human activity recognition \cite{sagar}. There are many examples where deep learning models have been developed for specific problems to demonstrate new methods and approaches \cite{example2}. Based on the above literature review, previous researchers have not investigated the use of an antenna as a sensor for picking up vibrations caused by a fault.

\begin{figure}[!t]
\centering
\subfloat[]{\includegraphics[width=1.5in]{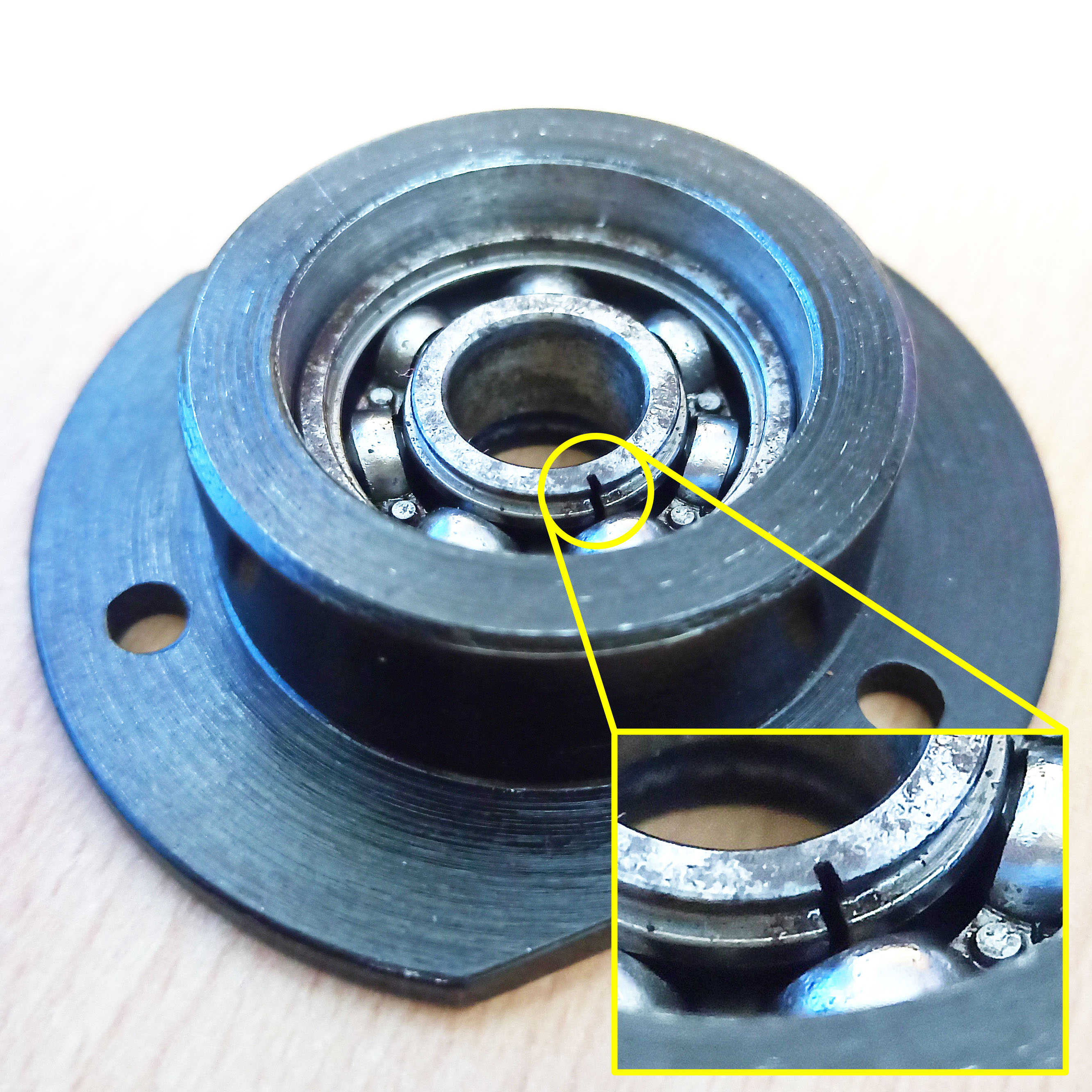}%
\label{fault_1}}
\hfill
\subfloat[]{\includegraphics[width=1.5in]{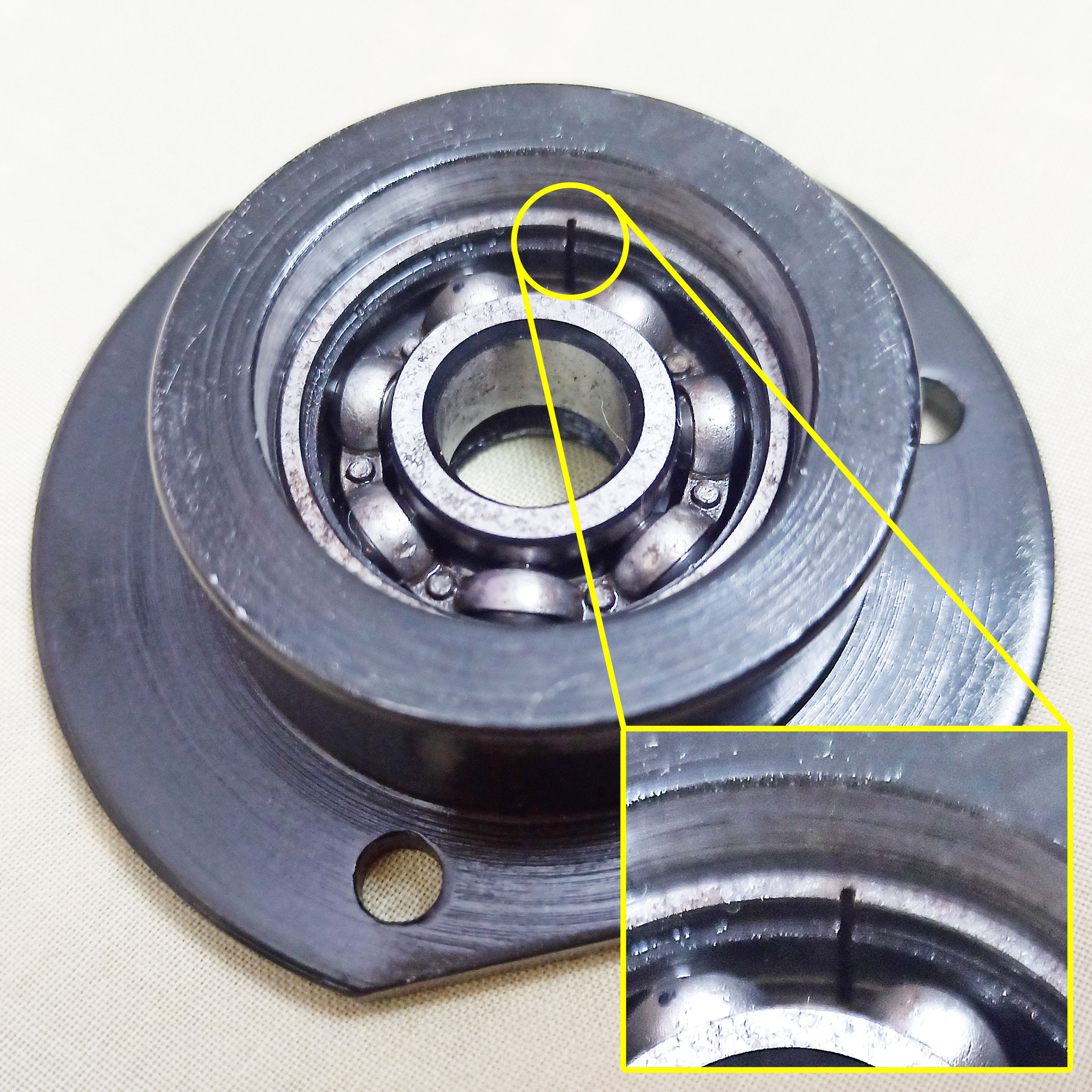}%
\label{fault_2}}\\
\subfloat[]{\includegraphics[width=1.7in]{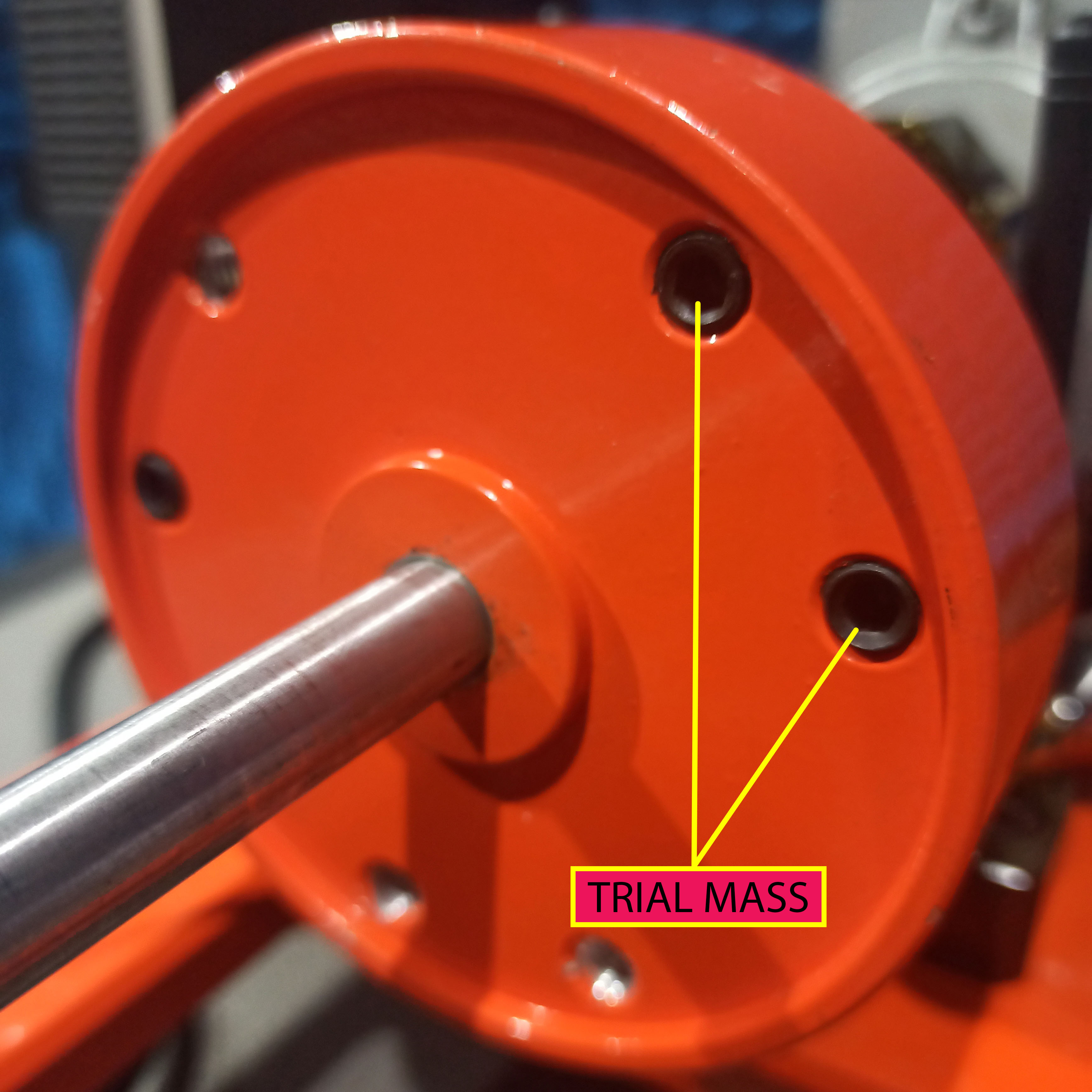}%
\label{fault_3}}
\caption{(a) Inner race faulty bearing (b) Outer race faulty bearing (c) Imbalance disk}
\label{fault}
\end{figure}

\begin{figure*}[!h]
\centering
\subfloat[]{\includegraphics[width=2.5in]{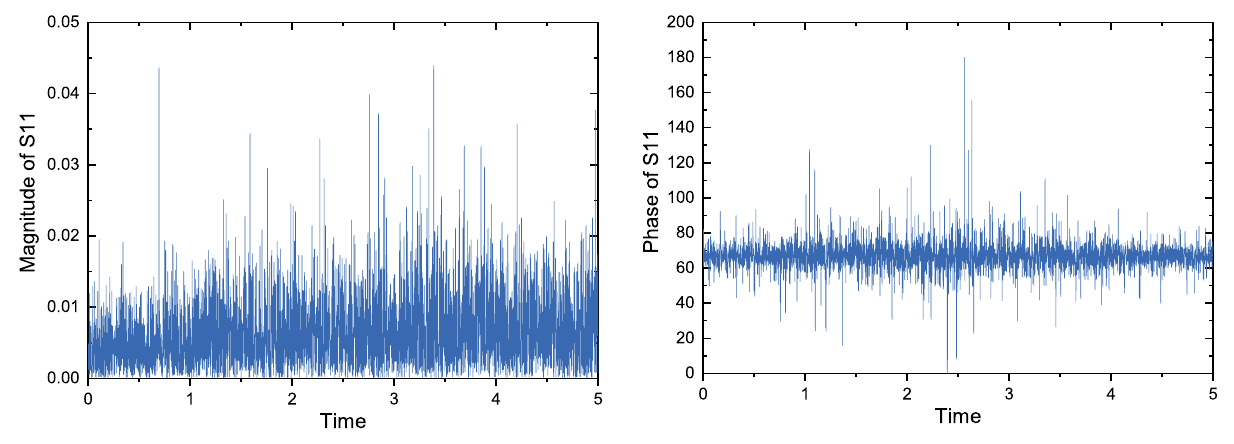}%
\label{raw1}}
\subfloat[]{\includegraphics[width=2.5in]{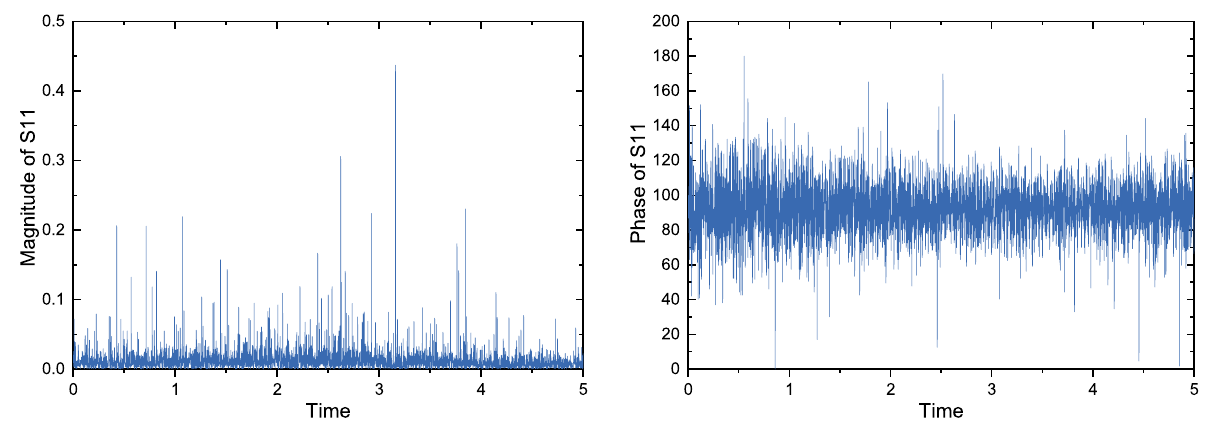}%
\label{raw2}}\\
\subfloat[]{\includegraphics[width=2.5in]{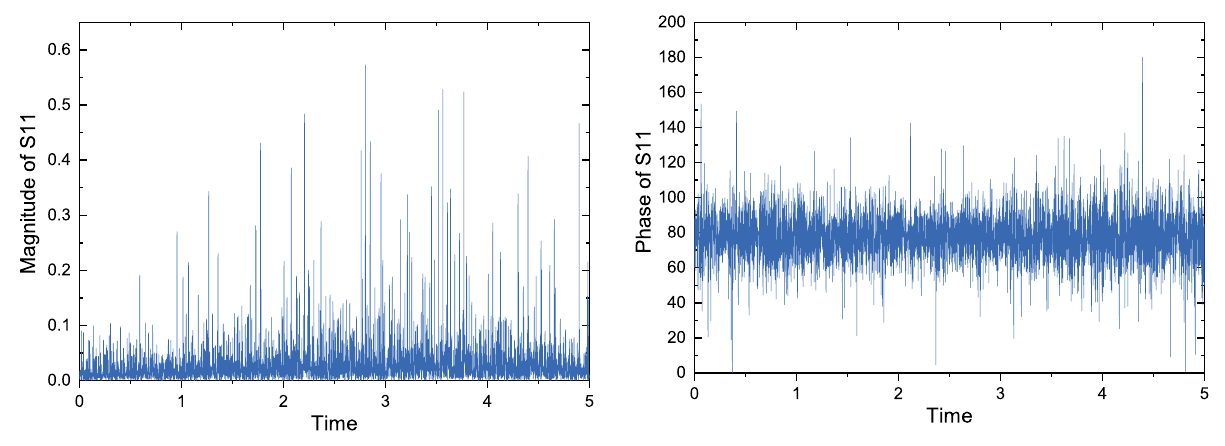}%
\label{raw3}}
\subfloat[]{\includegraphics[width=2.5in]{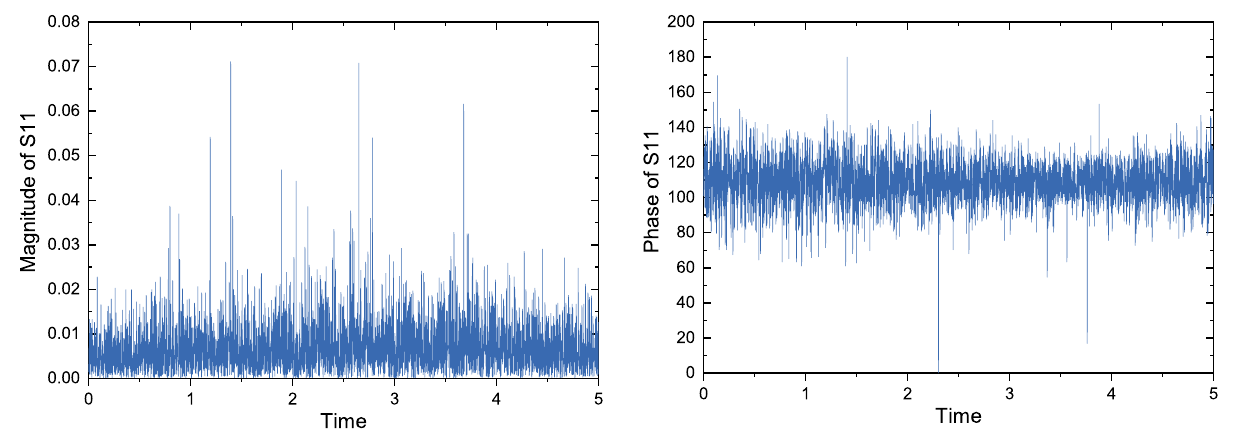}%
\label{raw4}}
\caption{Reflection coefficient for different operating conditions: (a) Normal (b) Imbalance (c) Inner race fault (d) Outer race fault}
\label{raw}
\end{figure*}

\begin{figure*}[!h]
    \centering
    \includegraphics[width=5in]{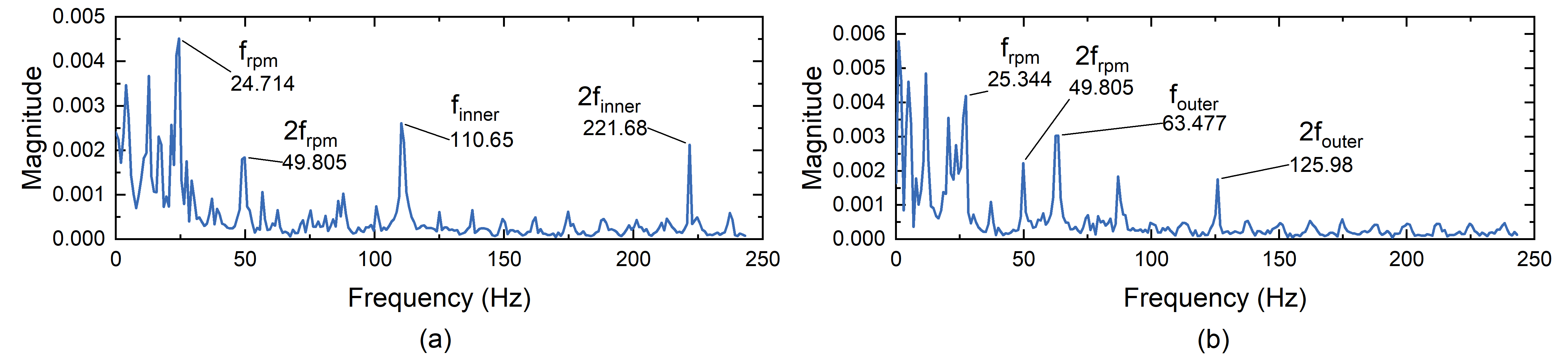}
    \caption{Characteristic vibration frequencies for (a) Inner Race Fault (b) Outer Race Fault }
    \label{bear}
\end{figure*}

\begin{figure}[!h]
\centering
\subfloat[]{\includegraphics[width=1.0in]{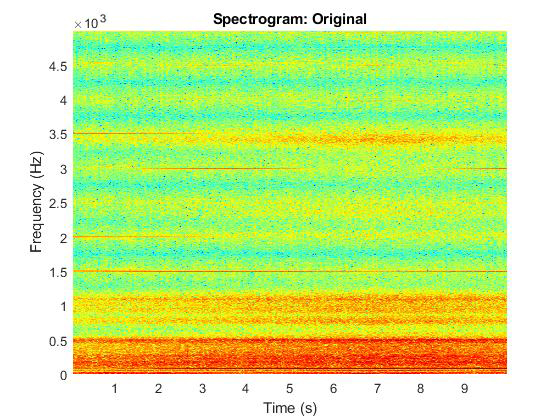}%
\label{specto1_mag}}
\subfloat[]{\includegraphics[width=1.0in]{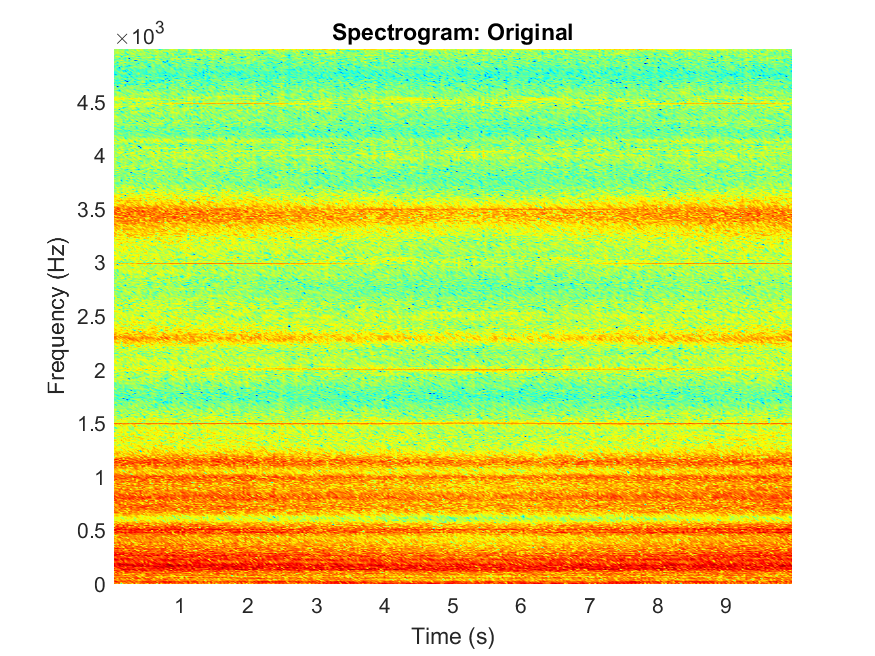}%
\label{specto2_mag}}\\
\subfloat[]{\includegraphics[width=1.0in]{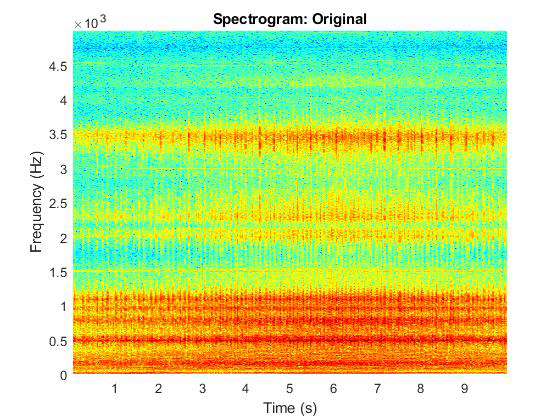}%
\label{specto3_mag}}
\subfloat[]{\includegraphics[width=1.0in]{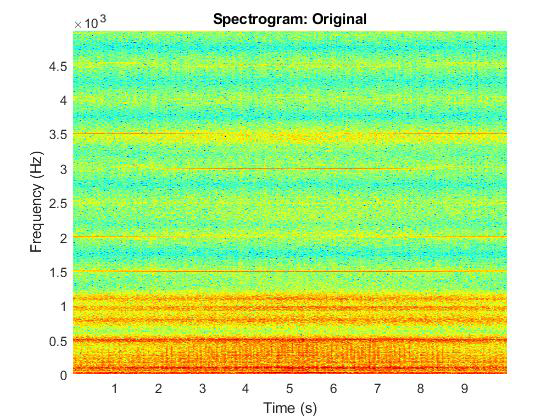}%
\label{specto4_mag}}
\caption{Spectrogram of S11 magnitude for different operating conditions: (a) Normal (b) Imbalance (c) Inner race fault (d) Outer race fault}
\label{specto_mag}
\end{figure}

\begin{figure}[!h]
\centering
\subfloat[]{\includegraphics[width=1.0in]{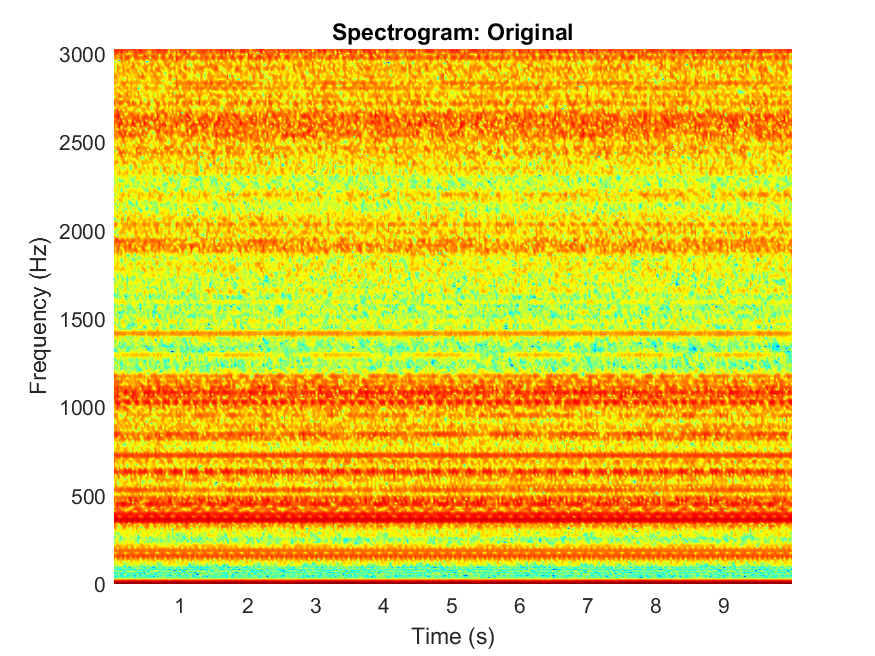}%
\label{specto1_phase}}
\subfloat[]{\includegraphics[width=1.0in]{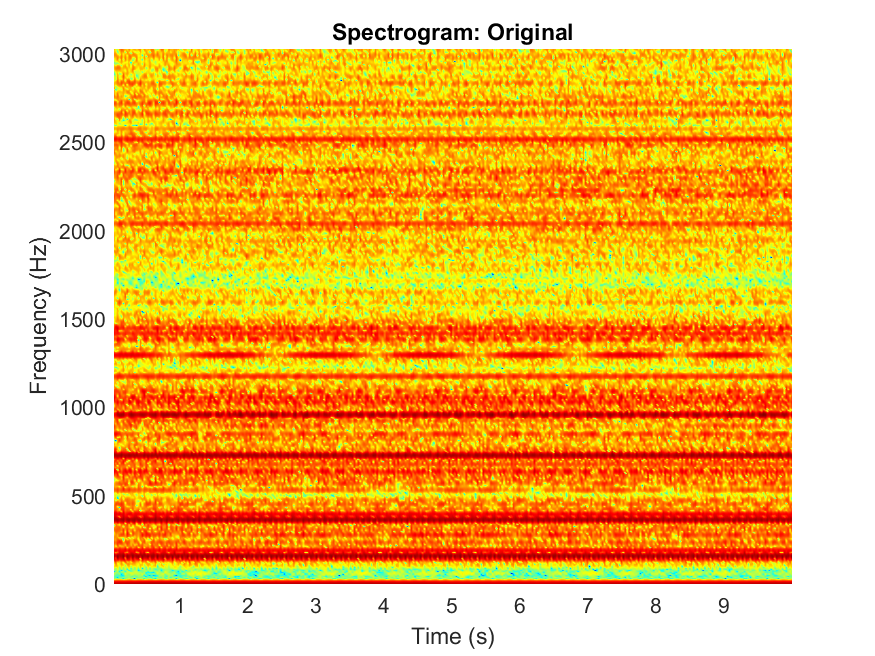}%
\label{specto2_phase}}\\
\subfloat[]{\includegraphics[width=1.0in]{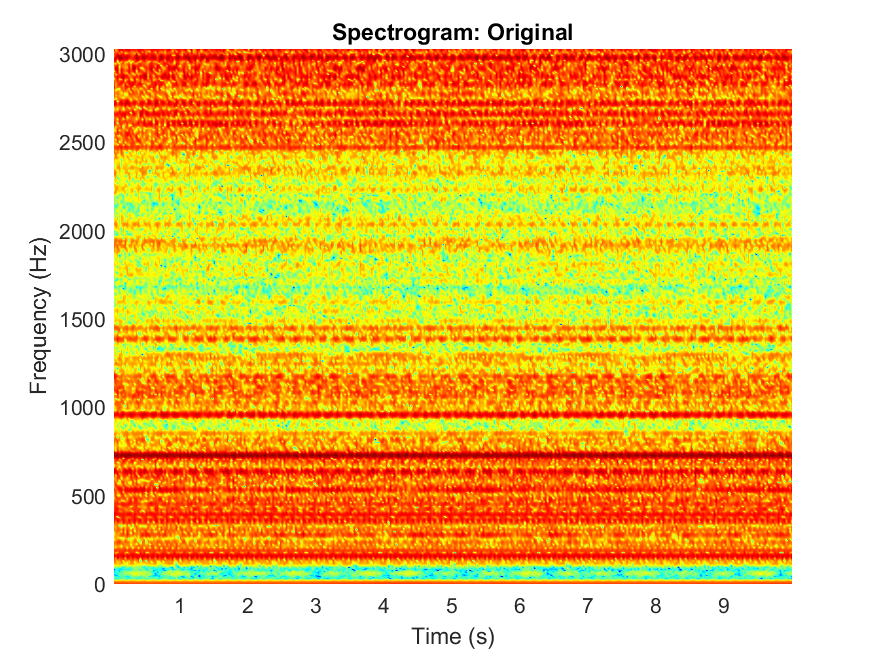}%
\label{specto3_phase}}
\subfloat[]{\includegraphics[width=1.0in]{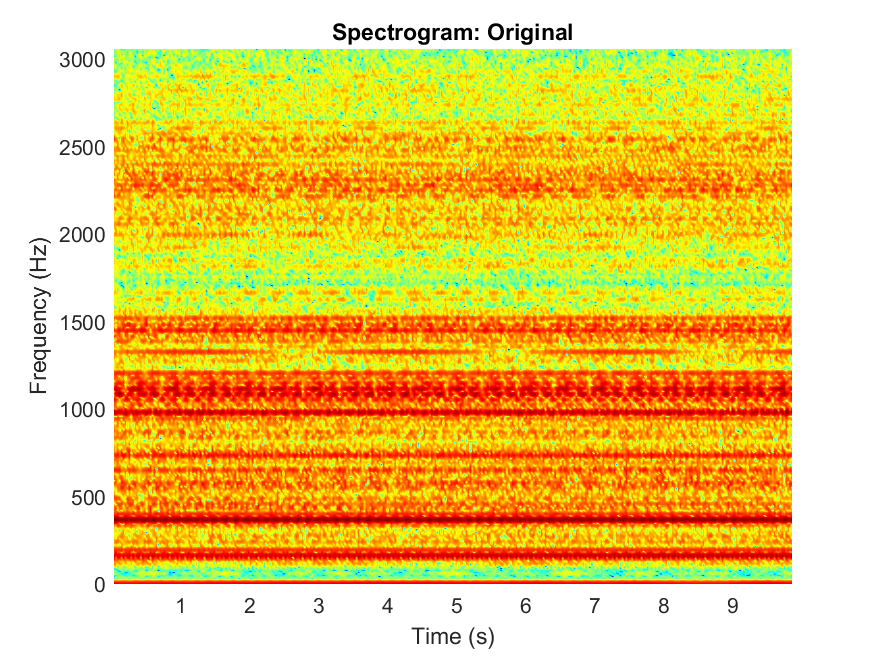}%
\label{specto4_phase}}
\caption{Spectrogram of S11 phase for different operating conditions: (a) Normal (b) Imbalance (c) Inner race fault (d) Outer race fault}
\label{specto_phase}
\end{figure}

In this context, the aim of this paper is to propose a non-contact, low-cost system to classify bearing faults and rotor imbalance of an induction motor using a single antenna's near-field effect. It appears that the site of vibration affects the near reactive field around the antenna which causes a change in the input impedance and the reflection coefficient of the antenna. The use of the antenna's reactive near-field effect for the classification of induction motor fault is not investigated in the literature until now. Though, \cite{sagar} explored the use of S-parameter for the classification of various human activities. The reflection coefficient ($S_{11}$) measurement is taken when the motor is at fault or imbalance considering near field effect. The vibrational disturbance particular to the fault is contained in the time-domain $S_{11}$ magnitude and phase, which is transformed into spectrogram to observe the unique characteristics related to the average energy at the fundamental band and the harmonics aroused due to specific fault. In the proposed work, the fault specific vibration frequencies of the induction motor captured by the antenna are experimentally evaluated and validated via FFT analysis of the S11 data for fault and imbalance conditions. For the classification of motor faults based on the unique time-frequency signatures, a DCNN model is designed and fine-tuned specifically for this investigation.

Many novel fault detection methods have been reported in the literature above. For example, article \cite{radar} has employed UWB RADAR to capture and process vibration signals to diagnose multiple bearing faults based on the signal energy at single fault frequency sub-band, \cite{flux} proposed a novel method to detect bearing fault based on stray flux measurements in various locations around the machine by analyzing the number of fault harmonics, and \cite{rotor} proposed a rotor speed-based bearing fault diagnosis method by principal component analysis on speed signal data. The literature review shows that past researchers have not exploited antenna near field behavior as a sensor to detect faults from vibrations. This study serves as one of the first attempts to employ antenna as a sensor to pick up vibrations due to bearing faults and imbalance in induction motors. To show the capability of our approach to integrate with the ongoing fourth industrial revolution and smart technology, we have implemented deep learning model which effectively classifies different faults without human intervention based on the magnitude and phase response of the reflection coefficient, and the results demonstrate its potential.

\section{Experimental Setup and reflection coefficient measurement}
The experimental setup of our study is shown in Fig. \ref{setup}. The test bench consists of a 3-phase AC induction motor with rating 220V, 50 Hz, coupled with a helical coupler to a shaft of diameter 12 mm and length 450 mm and operated at 1500 rpm (25 Hz). The shaft is fed through the bearing blocks which house the bearings. Two rotor disc measuring 100 mm in diameter and 30 mm in thickness is used to introduce imbalance by plugging metal screws in them. An antenna is placed near the bearing housing from where the vibrations are to be picked up as shown in Fig. \ref{setup_1b}. 

In order to analyze the fault conditions, the defects are artificially created. To conduct the defective bearing test, faults were introduced in the inner race and the outer race of HY-6201 bearings as shown in Fig. \ref{fault_1} and Fig. \ref{fault_2} respectively. The state of imbalance is present when the mechanical load of the induction motor is not evenly distributed, bringing the center of mass out of the motor shaft. The imbalance is introduced in the rotor shaft by adding trial masses on the rotor disc as shown in Fig. \ref{fault_3}.

When the bearing is operating, the local defect causes periodic peaks in the motors vibration signal. The frequency with which the peaks repeat is determined by the shaft speed, the bearing's physical parameters, and the defective bearing element. The vibration frequency due to inner and outer race defects is given by\cite{bearing_theory},
\begin{equation}
\label{eqn_inner}
f_{inner} = \frac{N_{b}}{2} f_{rpm}\left ( 1+\frac{D_{b}}{D_{c}}\cos \beta  \right )
\end{equation}
\begin{equation}
\label{eqn_outer}
f_{outer} = \frac{N_{b}}{2} f_{rpm}\left ( 1-\frac{D_{b}}{D_{c}}\cos \beta  \right )
\end{equation}
where $N_{b}$ is the number of balls, $f_{rpm}$ is the frequency of rotor, $\beta$, $D_{b}$ and $D_{p}$ are the contact angle of balls, ball diameter, and pitch diameter respectively. Furthermore, the vibrations caused by an imbalance in induction motors are a common problem. The imbalanced force that generates vibration can be expressed as,
\begin{equation}
F= m \times \vec r \times \omega^{2} \:[N]
\end{equation}
where $m$, $\vec r$ and $\omega$ denotes the unbalanced mass, distance of the unbalanced mass from center, and rotational speed respectively.

The measurements were taken at four operating conditions: normal condition, rotor imbalance condition, bearing with inner race fault condition and bearing with an outer race fault condition. The reflection coefficient $S_{11}$ of the antenna is measured using a vector network analyzer (VNA) under continuous-time mode for the four operating conditions. The measurements are also recorded at 3 different positions from the source of vibration (0 cm, 5 cm and 10 cm) to take into account the effect of distance on the classification. The measurement is performed inside an anechoic chamber. In total, 1440 datasets were collected (4 operating conditions $\times$ 3 frequencies $\times$ 3 test site $\times$ 40 trials). The length of the samples in the dataset is 5 seconds.

\begin{table}[!t]
\centering
\caption{Normalized average power for different operating conditions}
\label{power}
\resizebox{0.5\textwidth}{!}{%
\begin{tabular}{@{}lcccc@{}}
\toprule
                                        & \multicolumn{2}{c}{Magnitude}                                                                                                                                        & \multicolumn{2}{c}{Phase}                                                                                                                                            \\ \\\cmidrule(lr){2-3}\cmidrule(lr){4-5}
\multicolumn{1}{c}{Operating Condition} & \begin{tabular}[c]{@{}c@{}}Normalized \\ Average Power\end{tabular} & \begin{tabular}[c]{@{}c@{}}\% increase in power\\ compared to normal \\ condition\end{tabular} & \begin{tabular}[c]{@{}c@{}}Normalized \\ Average Power\end{tabular} & \begin{tabular}[c]{@{}c@{}}\% increase in power\\ compared to normal \\ condition\end{tabular} \\ \hline \\
Normal                                  & 0.00002895                                                          & -                                                                                              & 0.00002542                                                          & -                                                                                              \\ \\
Imbalance                               & 0.00003121                                                          & 7.80 \%                                                                                        & 0.00002677                                                          & 5.31\%                                                                                         \\\\
Inner Race Fault                        & 0.00003027                                                          & 4.55 \%                                                                                        & 0.00002617                                                          & 2.95\%                                                                                         \\\\
Outer Race Fault                        & 0.00002958                                                          & 2.17 \%                                                                                        & 0.00002574                                                          & 1.25\%                                                                                         \\ \bottomrule
\end{tabular}
}
\end{table}

Equation \ref{eqn1} shows the reactive near field region around the antenna where the energy of a certain amount is retained in the reactive field. 
\begin{equation}
    Reactive\ near\ field < 0.62\sqrt{\frac{L^{3}}{\lambda }}
    \label{eqn1}
\end{equation}
where L is the largest dimension of the antenna and $\lambda$ is the wavelength. When an external conductor comes within this region, there is a transfer of field energy to the electrons in the conductor which results in the loss of energy in the antenna. This effect causes a shift in the antenna impedance. In our case, the vibration causes a varying shift in the impedance of the antenna with respect to time. This also results in the change in reflection coefficient as per Equation \ref{eqn2}. It is the ratio of the reflected signal voltage to the incident signal voltage. A detailed discussion on the related theory can be referred here\cite{balani}.
\begin{equation}
    \textnormal{Reflection Coefficient}, \Gamma= \frac{V_{reflected}}{V_{incident}} = \frac{Z_{i}-Z_{o}}{Z_{i}+Z_{o}}
    \label{eqn2}
\end{equation}
where $Z_{i}$ and $Z_{o}$ are the antenna's input impedance and output impedance respectively.

Fig. \ref{raw} shows the measured reflection coefficient (magnitude and phase) for different operating conditions. It is observed from Fig. \ref{raw} that for faulty condition the magnitude of $S_{11}$ signal is significantly greater due to near field effect. We have analyzed the average power content of the signals for the four operating conditions shown in Fig. \ref{raw}. The average power of a discrete signal x(n) can be expressed as,
\begin{equation}
P_{avg}= \frac{1}{N} \sum_{n=1}^{N} \left | x(n) \right |^{2}
\end{equation}

\begin{figure*}[!t]
    \centering
    \includegraphics[width=0.6\textwidth]{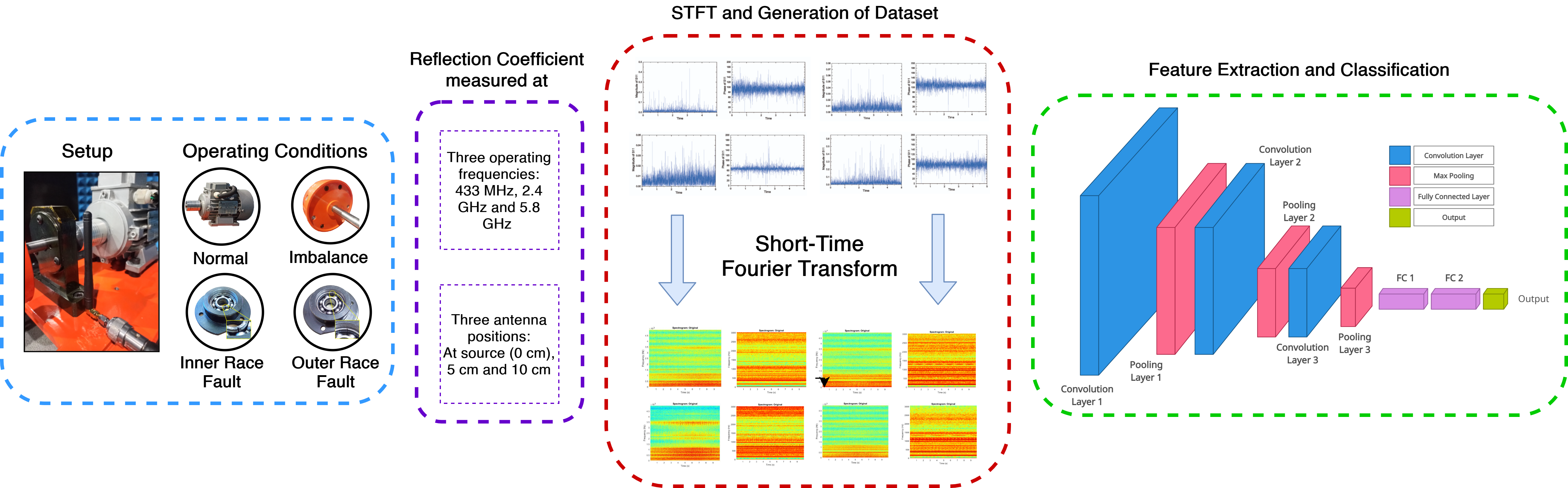}
    \caption{Schematic representation of approach and DCNN model}
    \label{dcnn}
\end{figure*}

The normalized average power of magnitude and phase signal for the different operating conditions is shown in Table \ref{power}. It is observed for magnitude data that, when an imbalance fault occurs, the average power changes by 7.8\% when compared to normal condition. With inner race fault and outer race fault the average power changes by 4.55\% and 2.17\% respectively. Similarly, for phase data, with imbalance average power changes by 5.31\%, with inner race fault and outer race fault the average power changes by 2.95\% and 1.25\% respectively. 

It is observed from Table \ref{power} that the magnitude and phase response of the antenna's reflection coefficient for the four operating conditions is different from each other which implies that the antenna can be successfully utilized as a sensor to detect vibrations due to various faults. This is due to the fact that fault-specific vibration creates varying levels of antenna impedance mismatch and because of this impedance mismatch, a portion of EM wave is reflected back and not transmitted by the antenna. The reflected wave has its own magnitude and phase characteristics. Thus, the fault-specific vibration results in unique reflection coefficient magnitude and phase response over time. In comparison to the signal power analysis done in \cite{radar} to distinguish multiple faults using RADAR, our approach also indicates that different faults have varied power content.

Furthermore, we have also performed spectral analysis on the S11 data to verify the existence of characteristic vibration frequencies related to the inner/outer race faults. In our experiment, we have used HY-6201 bearing where $N_{b}=7$, $\beta = 0\degree$, $D_{b}= 6 mm$ and $D_{p}= 22 mm$. Using equations (\ref{eqn_inner}) and (\ref{eqn_outer}), the theoretical values of $f_{inner}$ and $f_{outer}$ is found to be 111.3 Hz and 63.8 Hz respectively. As seen in Fig. \ref{bear}, the presence of discrete peaks at $f_{inner}=110.65\:Hz$ and $f_{outer}= 63.47\:Hz$ along with its harmonics clearly indicates the presence of a fault in the inner and outer races. This further validates the feasibility of using an antenna as a sensor for fault detection. In comparison to the spectral analysis performed in \cite{mem} to compare the theoretical and measured discrete fault frequencies, our approach also shows that it can detect the fault frequencies that match the theoretical value.

Next, the Short-Time Fourier Transform (STFT) is then applied to generate the spectrogram of the measured reflection coefficient data. Spectrograms represent the signal intensity of a signal at various frequencies over time. The sampled data is transformed by STFT as follows for each segment of the signal,
\begin{equation}
\begin{split}
X(m,\omega) = \sum_{n=-\infty}^{\infty} x(n)w(n-m)e^{-j\omega n}
\end{split}
\end{equation}
where $x(n)$ and $w(m)$ is the signal to be transformed and window function respectively. The spectrogram is obtained by taking the magnitude squared of $X(m,\omega)$. The spectrogram of $S_{11}$ magnitude and phase for the four operating conditions are shown in Fig. \ref{specto_mag} and Fig. \ref{specto_phase} respectively. It is observed that the time-varying signals for different operating conditions has unique characteristics due to the fault-specific vibrations induced. In the spectrogram, we can observe the different frequency components with magnitude depicted by color for each operating condition. This implies the possibility of pattern recognition using deep learning model.

\section{Classification Technique}

With the development of machine learning, deep learning has emerged as an effective way for the diagnosis of faults. Compared to conventional machine learning approaches, DL has achieved good results, but the use of DL for fault diagnosis is still developing. Convolutional neural network has also been extended to fault detection as one of the most successful DLs.
\begin{figure*}[!h]
\centering
\subfloat[]{\includegraphics[width=1.5in]{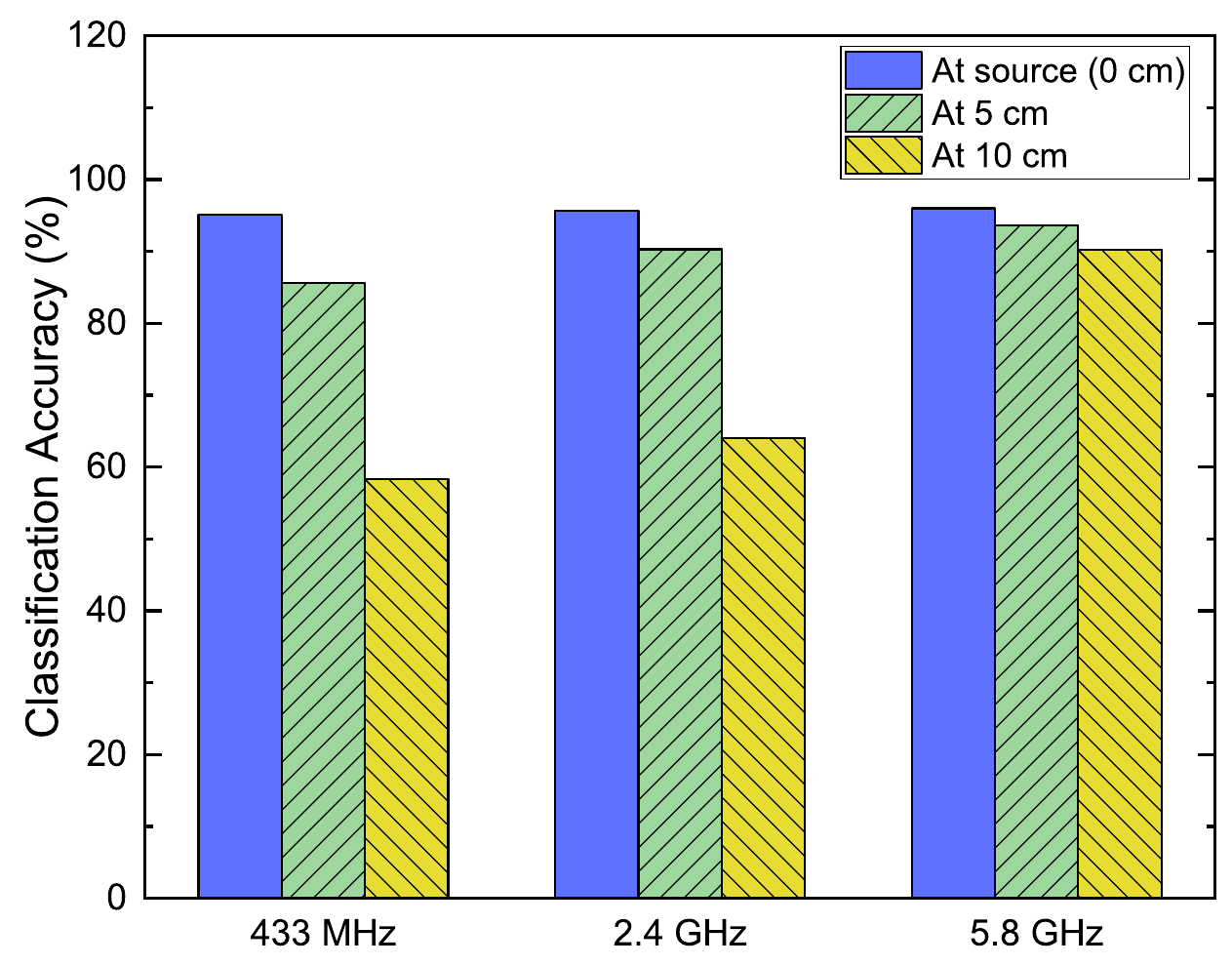}%
\label{accuracy_1}}
\subfloat[]{\includegraphics[width=1.5in]{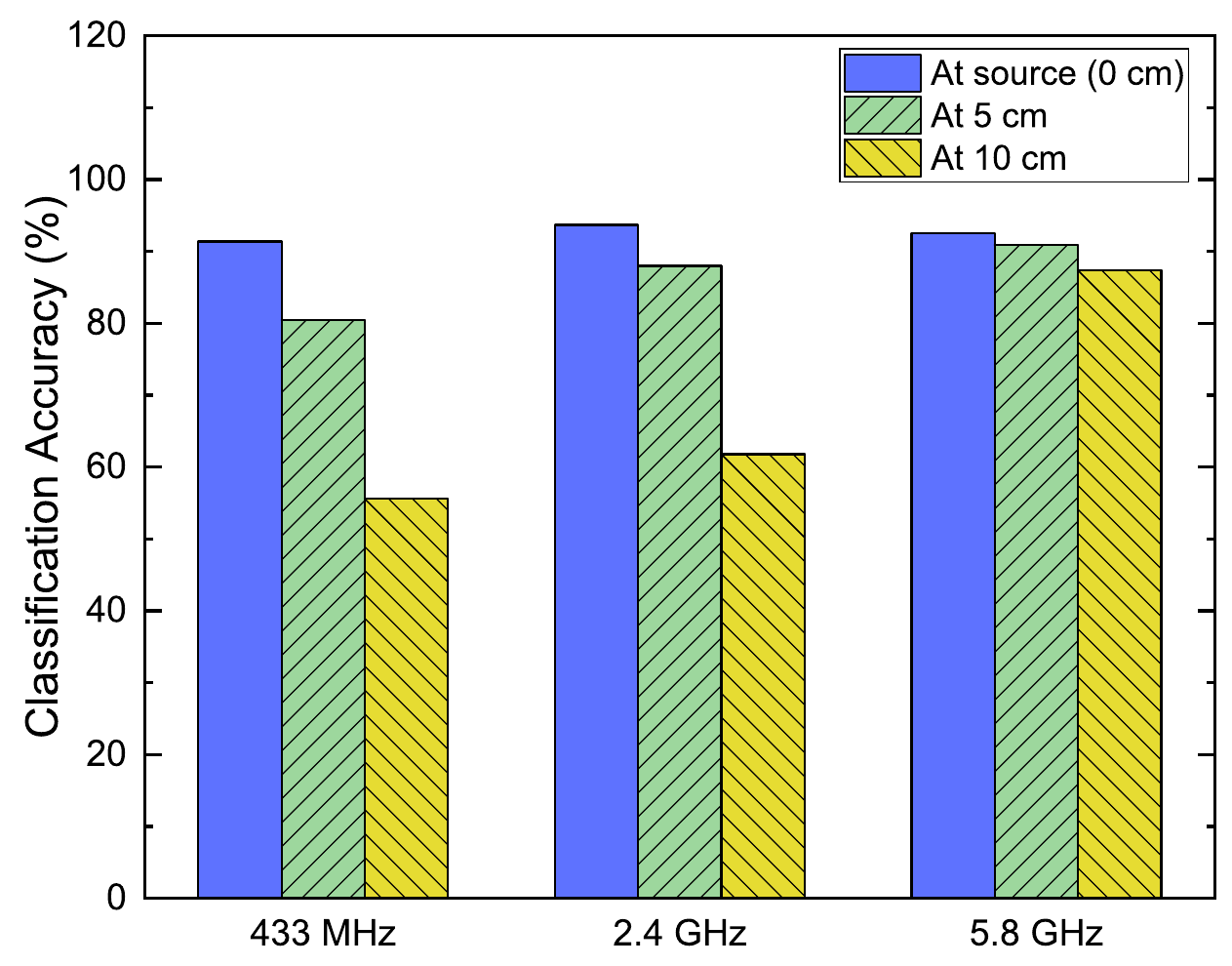}%
\label{accuracy_2}}
\subfloat[]{\includegraphics[width=1.5in]{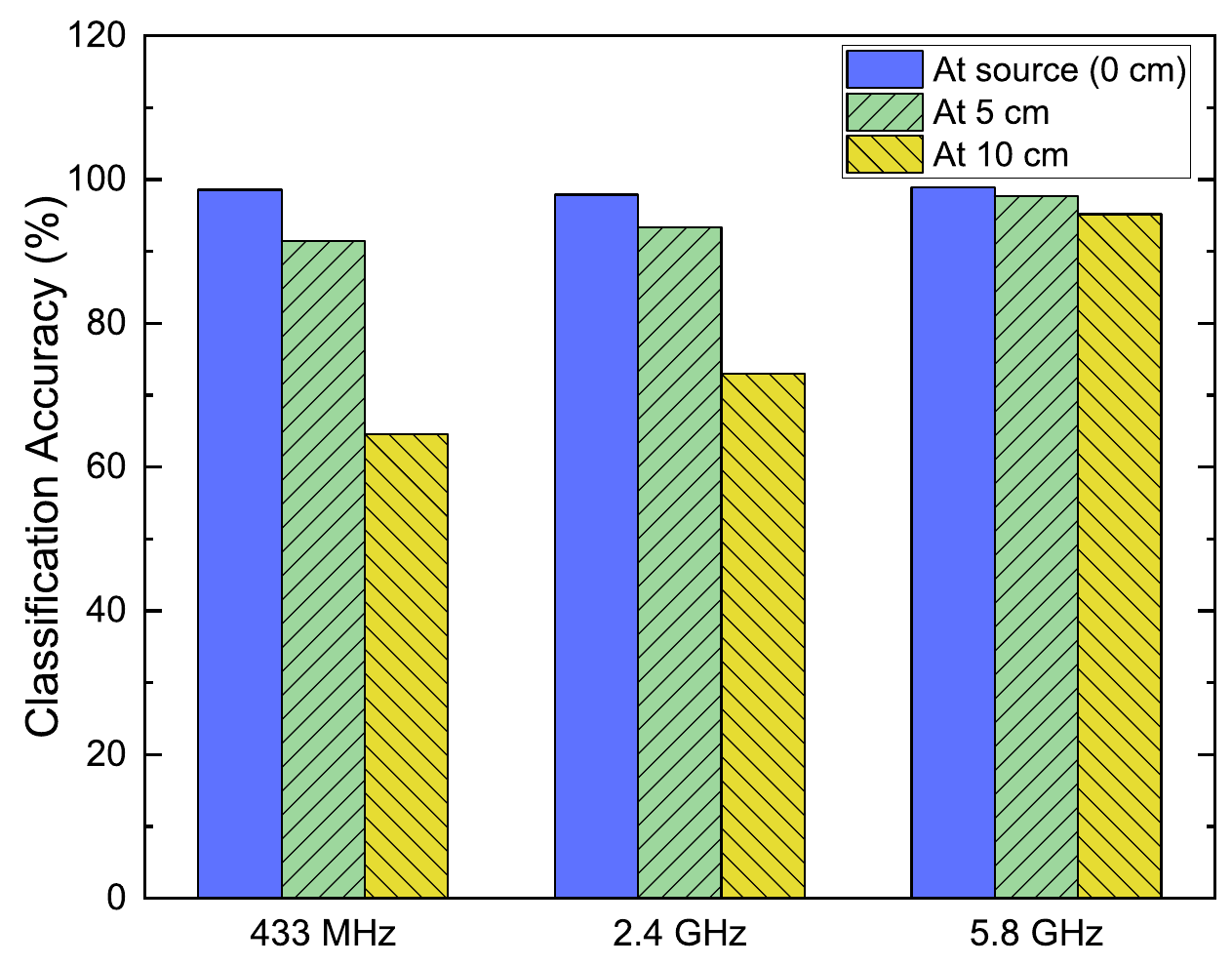}%
\label{accuracy_3}}
\caption{Classification accuracy using (a) $S_{11}$ Magnitude (b) $S_{11}$ Phase and (c) Combination of magnitude and phase}
\label{accuracy}
\end{figure*}
Deep convolutional neural network exhibits superior performance in terms of both computational complexity and classification accuracy. The key advantage of this deep learning approach is the ability to learn non-linear representations of the raw signal to a greater degree of abstraction and sophistication. This is significant from a technical point of view because covariates often have no linear impact on the result of the fault diagnosis. In our study, we have implemented DCNN due to its excellent empirical performance in domains like image processing and speech processing. Furthermore, it has been observed that using spectrograms of S-parameters with the DCNN produces better results than using the time-domain technique. The deep learning approach is ideal because the spectrogram data we plan to classify can be considered as an image.

A DCNN is made up of a number of convolutional and pooling layers, which are followed by one or more fully linked layers. A sequence of filters or kernels is usually found in each convolutional layer. The convolutional procedure can be performed across the full signal by sliding the filter along it. A feature map is the product of this technique. The convolution operation can be expressed as,
\begin{equation}
C_{l}^{n} = ReLU\left ( \sum_{m} v_{l-1}^{m} \ast w_{l}^{n} + b_{l}^{n} \right )
\end{equation}
where $C_{l}^{n}$ is the output of the nth filter in convolutional layer l; ReLU denotes the nonlinear activation function; $w_{l}^{n}$ denotes the current layer's nth filter; $v_{l-1}^{m}$ is the previous layer's mth output; $\ast$ represents the convolution and $b_{l}^{n}$ is the bias. To reduce the spatial dimensions of the feature map and avoid over-fitting, a pooling layer is often added to the convolutional layer. Max pooling is the most prevalent type of pooling, which only takes the input's most significant component (the highest value). The max pooling is defined as,
\begin{equation}
P_{r+1}^{n} = max \; C_{r}^{n} (t)
\end{equation}
where $P_{r+1}^{n}$ and $C_{r}^{n}$ denotes the output of the pooling layer and the feature map respectively. With the features created after consecutive convolution and pooling, the fully-connected layer combined with a softmax function is finally used for classification. The output layer is given by,
\begin{equation}
y = S(Wa+b_{c})
\end{equation}
where $S()$ denotes the activation function, which is a softmax function for classification problems, $a$ and $W$ denotes the neurons and weight matrix respectively in the fully-connected layer connected to the output layer and $b_{c}$ denotes the bias. 

In our study, we have converted the time domain raw signals into a time-frequency domain using STFT, which provides a way to explore the two dimensional features of the raw signal. Once the raw signals have been converted to spectrogram images, a DCNN is trained to classify the images relevant to specific operating conditions. The size of each spectrogram is set to 100$\times$100 and is normalized between 0 to 1. Fig. \ref{dcnn} shows the schematic representation of our approach and architecture of our DCNN model. The DCNN consists of 3 convolutional layers each followed by pooling layers. A fully connected layer acts as the final stage for the classification. The number of convolutional filters used in the first, second and third layers is 64, 64 and 128 respectively. The convolutional filter size is set to 3$\times$3 for all the layers and the reduction ratio for the pooling layer is 2:1. The pooling layer is followed by two fully connected layers and a dropout rate of 0.5 is employed to avoid over-fitting. These hyper-parameters were described by a heuristic search. The input of this DCNN is the spectrogram images of S11 magnitude and phase of size 100$\times$100. Multiple convolution filters are used in the convolutional layer to obtain the feature maps of the input spectrograms. The spectrogram feature maps for different operating conditions contain unique information because they contain different characteristic fault frequencies and varying power content. The max pooling layer then reduces the dimension of the spectrogram feature maps. The fully-connected layers are used to compute the class scores after the set of three consecutive convolutional and pooling layers. For the construction and the training of the DCNN, we have used the deep learning toolbox in MATLAB. As the training process completes and the new unforeseen spectrogram image is fed to the DCNN input, it generates the output corresponding to the class of the spectrogram. For the training process, we have used a 5-fold cross-validation method where 20\% of the training data is used for validation.

\begin{figure*}[!h]
\centering
\subfloat[]{\includegraphics[width=1.5in]{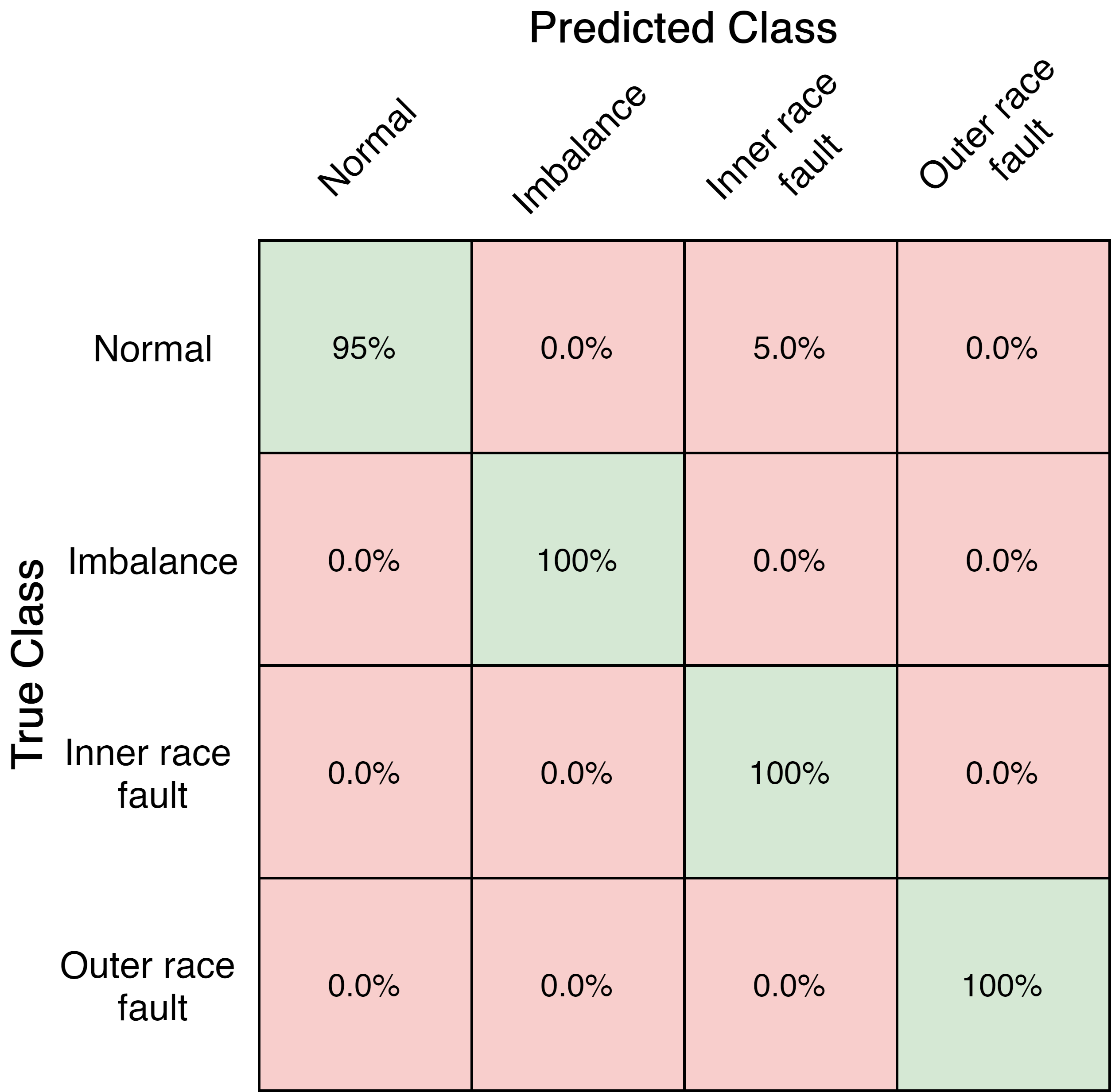}%
\label{confusion_1}}
\subfloat[]{\includegraphics[width=1.5in]{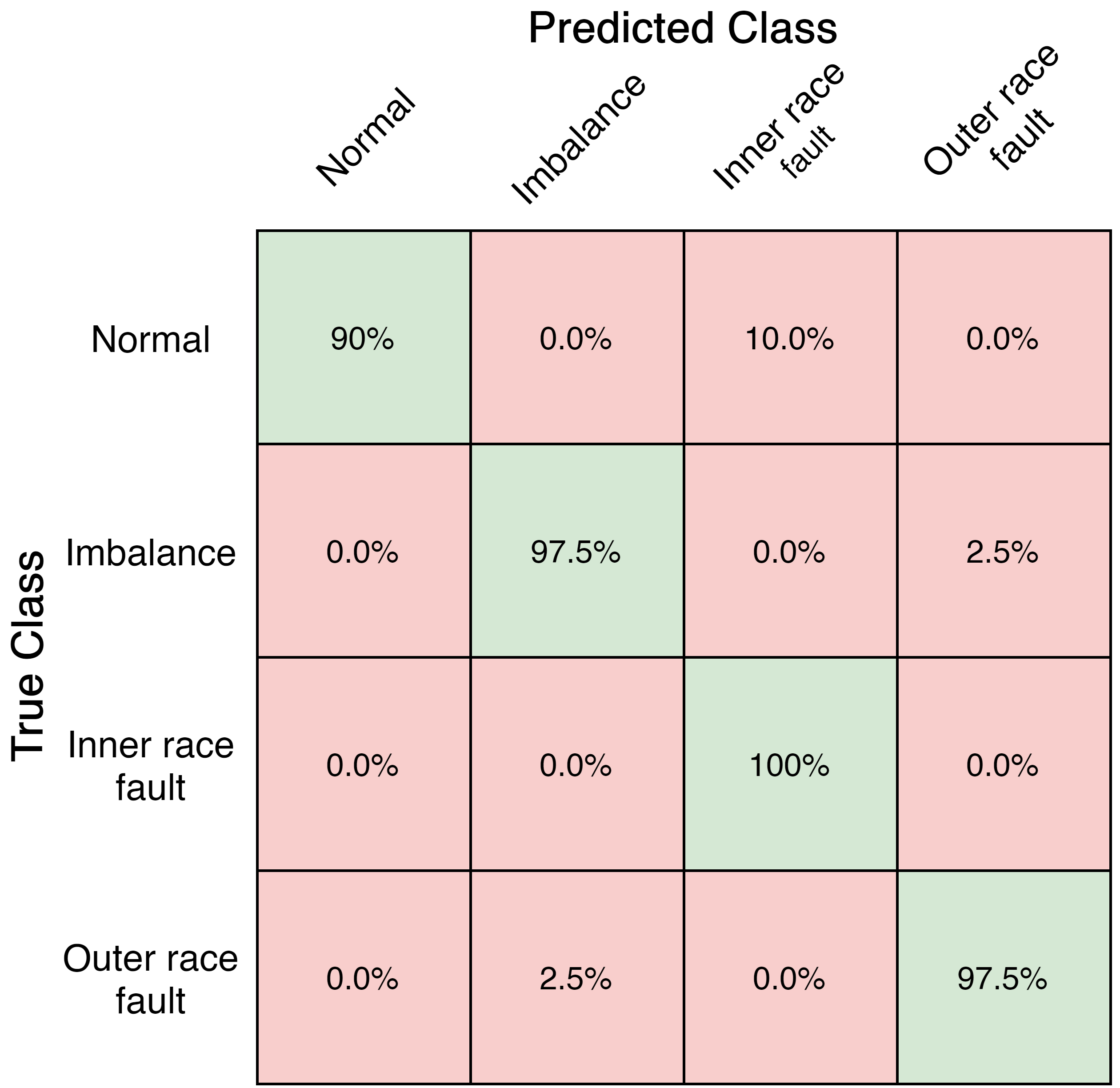}%
\label{confusion_2}}
\subfloat[]{\includegraphics[width=1.5in]{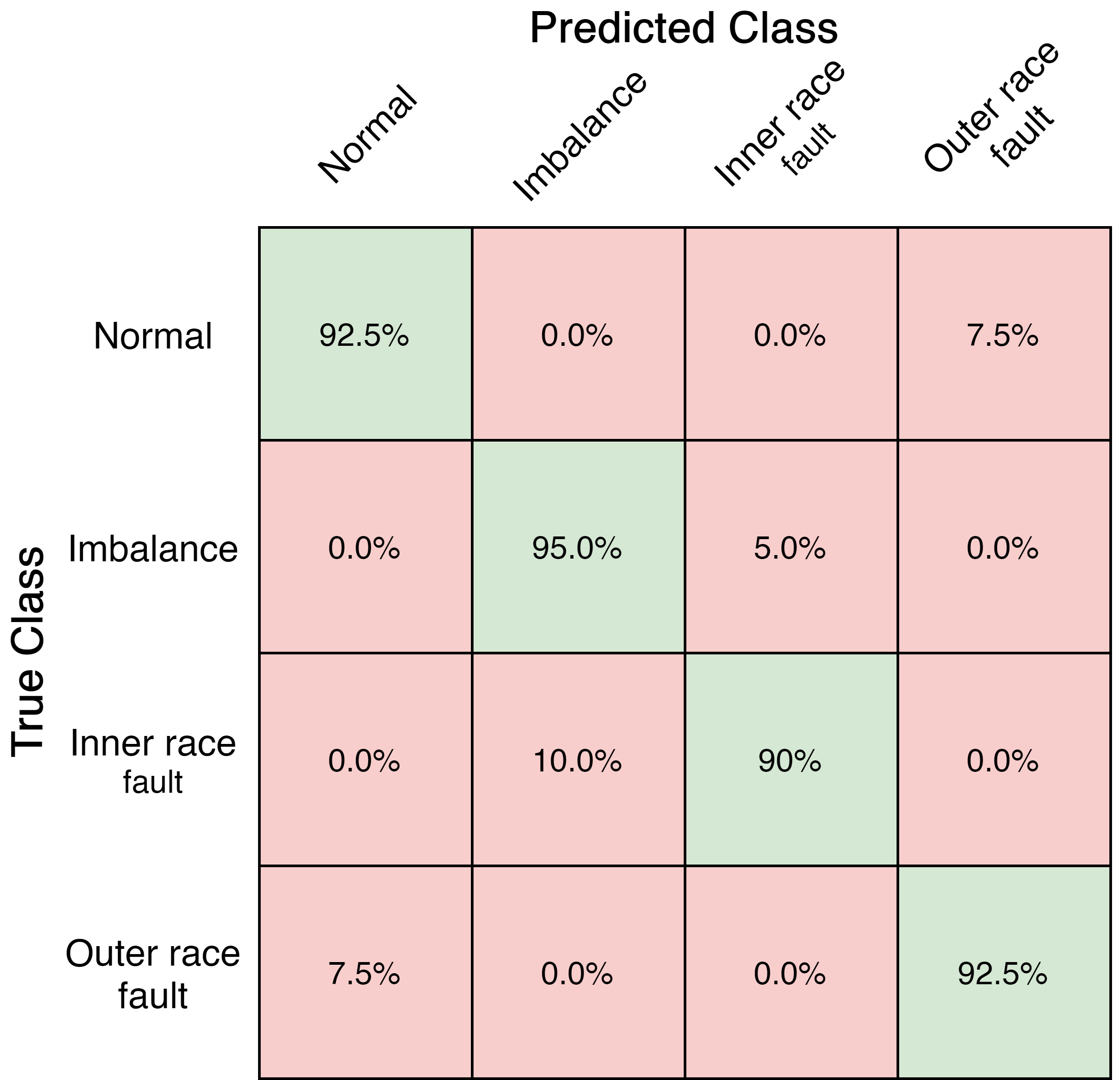}%
\label{confusion_3}}
\caption{Confusion matrix for (a) $S_{11}$ Magnitude and phase (b) $S_{11}$ Magnitude and (c) $S_{11}$ phase}
\label{confusion}
\end{figure*}

\section{Classification Results}

In this section, the classification accuracy for each of the four conditions are investigated. Fig. \ref{accuracy_1}, b, and c shows the classification accuracy at the three operating frequencies based on a) magnitude, b) phase, and c) both magnitude and phase of the measured reflection coefficient, respectively at three different antenna positions. Fig. \ref{accuracy_1} shows that the antenna placed at the source produced similar classification accuracy (96\%) for the three frequencies. A decrease in accuracy is observed when the antenna is placed at 5 cm and 10 cm, at 433 MHz and 2.4 GHz. However, the decrease in accuracy at 5.8 GHz is very less compared to the other frequencies. On the other hand, Fig. \ref{accuracy_2} shows a similar trend as Fig. \ref{accuracy_1} which indicates that the $S_{11}$ phase also contains unique information regarding the fault conditions. However, when the magnitude and phase is combined, we were able to achieve a higher classification accuracy compared to using magnitude and phase alone, as shown in Fig. \ref{accuracy_3}.

It is observed that as the antenna moves away from the vibration source, there is a decrease in the classification accuracy for the three frequencies. However, the decrease in the accuracy is prominent in the case of 433 MHz and 2.4 GHz as compared to 5.8 GHz. This is due to the fact that the range of the reactive near field of the antenna directly varies as a function of the frequency. Moreover, it is observed that both magnitude and phase data at 5.8 GHz, provide the best classification accuracy for the three antenna positions. This result is likely because the range of the near reactive field at other frequencies is smaller than that at 5.8 GHz. So the near reactive field is not disrupted by the vibrations when the antenna is moved away due to their small range. For the three antenna positions, the highest accuracy is achieved at 5.8 GHz which is more than 95\%.

\begin{figure}[!b]
    \centering
    \includegraphics[width=2.5 in]{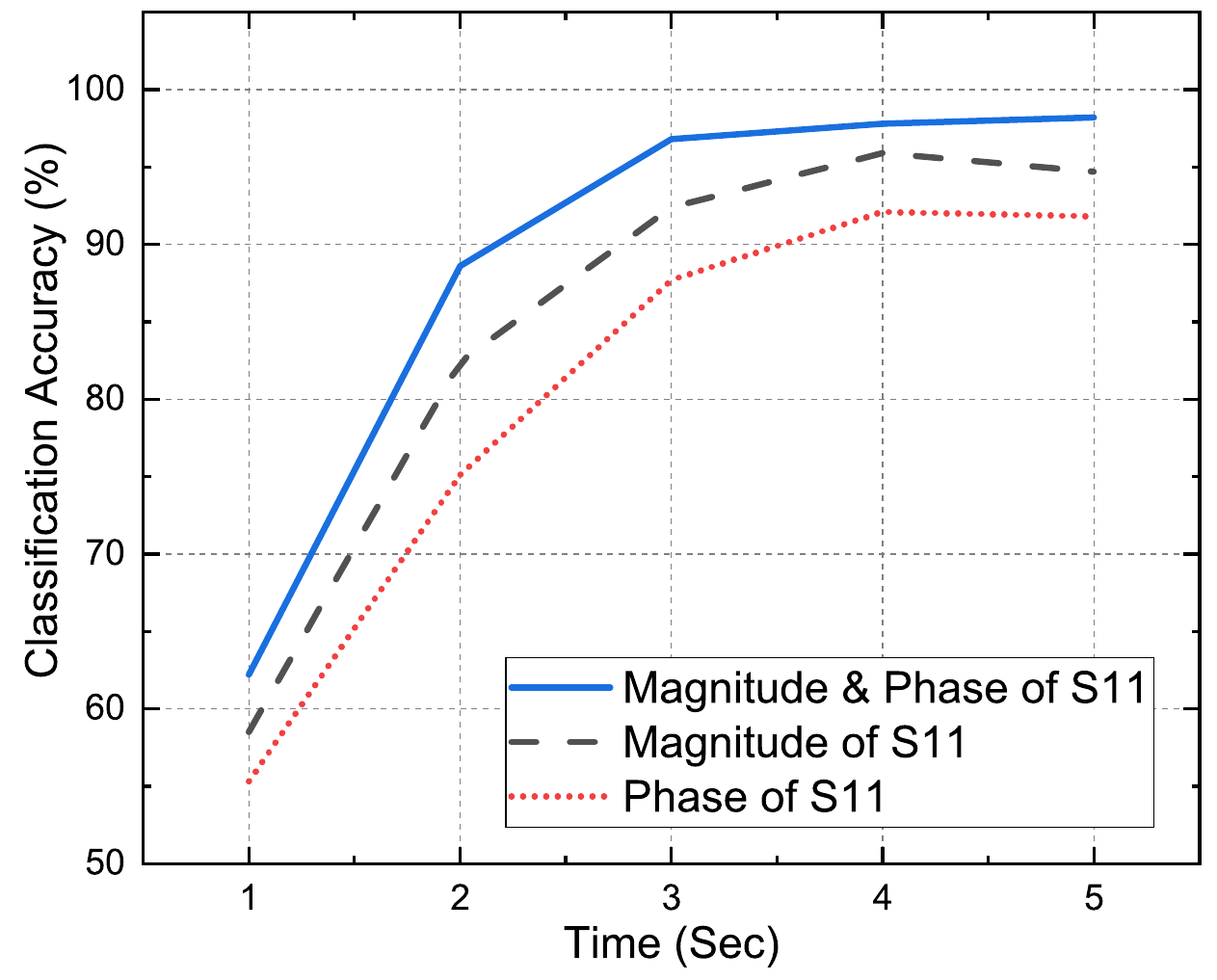}
    \caption{Classification accuracy vs. time window}
    \label{window}
\end{figure}

\begin{table}[!b]
\centering
\caption{Performance metrics} 
\label{metric} 
\begin{tabular}{@{}lcccc@{}}
\toprule
\multicolumn{1}{c}{} & Accuracy & Sensitivity & Precision & F1 Score \\ \midrule
Magnitude and Phase  & 98.2 \%  & 98.75       & 98.75     & 98.75    \\ \\
Magnitude            & 96.0 \%  & 96          & 96.25     & 96       \\ \\
Phase                & 92.1 \%  & 92.75       & 92.75     & 92.75    \\ \bottomrule
\end{tabular}
\end{table}

The effect of the time duration of the signal on the classification accuracy is also investigated. Fig. \ref{window} shows the classification accuracy versus the time duration of the signal at an operating frequency of 5.8 GHz with antenna at the origin. It is seen that the classification accuracy increases with the time duration of the signal. A highest value of 98.2\% is achieved using both magnitude and phase of $S_{11}$ with 3 second time window. An accuracy of 96.0\% and 92.1\% is achieved using magnitude and phase of $S_{11}$ respectively with a time window of 4 seconds. The performance metrics for the three cases are shown in Table. \ref{metric}.

\begin{figure}[!t]
    \centering
    \includegraphics[width=2.5 in]{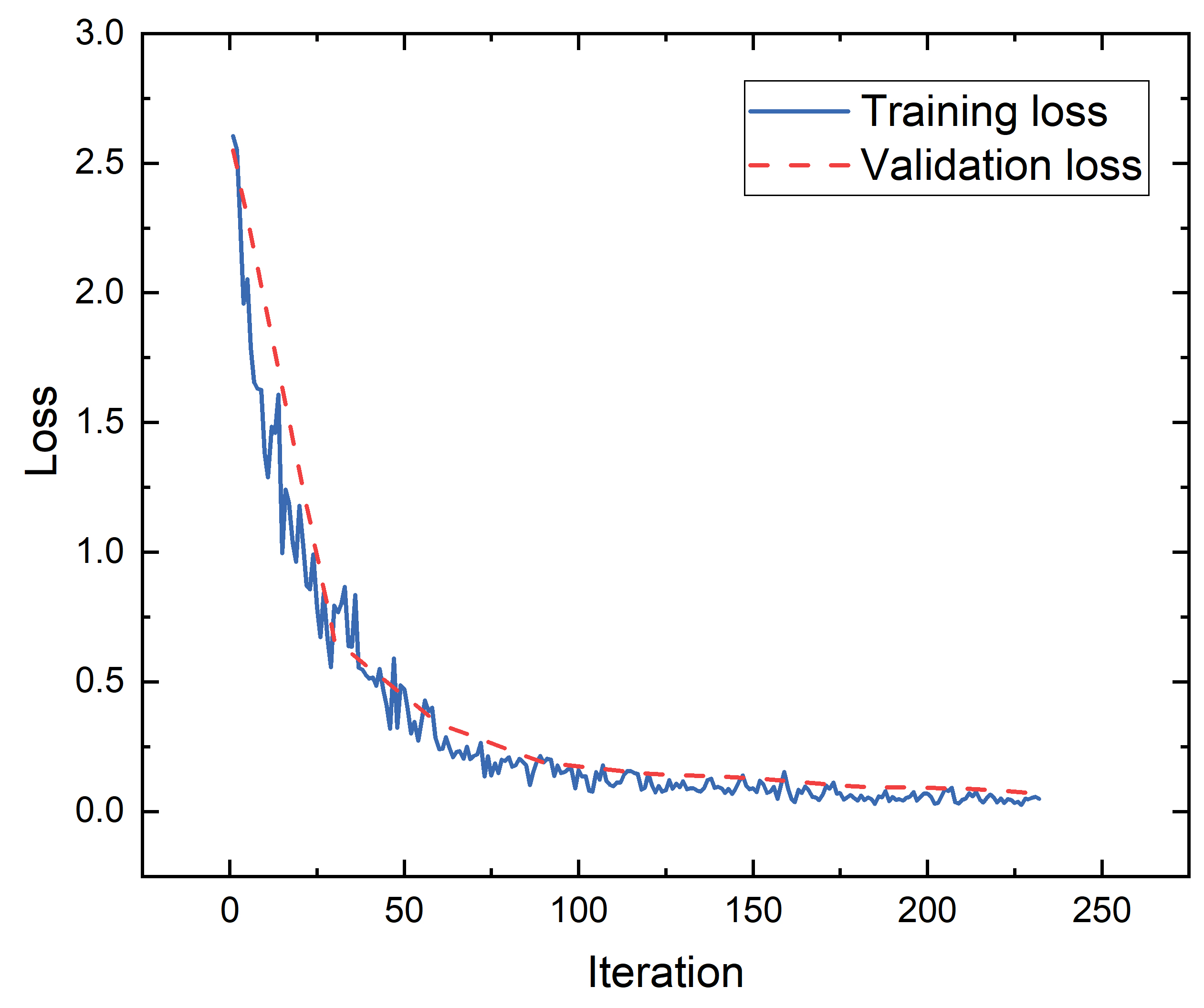}
    \caption{Model's learning curve}
    \label{loss}
\end{figure}

To investigate the performance of the classifiers, the confusion matrix for the three cases is shown in Fig. \ref{confusion}. From the confusion matrix, it is observed that the normal condition is confused with the inner race fault condition in Fig. \ref{confusion_1}. In Fig. \ref{confusion_2} the confused classes are normal condition with inner race fault, imbalance with outer race fault and vice-versa. The confused classes for Fig. \ref{confusion_3} are normal condition with outer race fault, imbalance with inner race fault, inner race fault with imbalance and outer race fault with normal condition. Furthermore, Fig. \ref{loss} shows the learning curve of our model. It is seen that the training loss and validation loss decreases and converges to a point of stability, and the generalization gap is minimal. We can conclude that the model is able to generalize different fault-specific signal patterns optimally.

\begin{table*}[!t]
\centering
\caption{Comparison of proposed method with existing methods}
\resizebox{0.8\textwidth}{!}{%
\begin{tabular}{@{}lccccccc@{}}
\toprule
\multicolumn{1}{c}{Method}                                                                         & \begin{tabular}[c]{@{}c@{}}Faults identified/ \\ Vibration source\end{tabular}             & \begin{tabular}[c]{@{}c@{}}Measured \\ Signal\end{tabular}                & \begin{tabular}[c]{@{}c@{}}Sensor\\ Used\end{tabular}                                         & \multicolumn{1}{l}{\begin{tabular}[c]{@{}l@{}}Complexity \\ of Sensors\end{tabular}} & \begin{tabular}[c]{@{}c@{}}Method of \\ Signal Analysis\end{tabular}                           & \begin{tabular}[c]{@{}c@{}}Signal Duration\\ Required (Sec)\end{tabular} & \begin{tabular}[c]{@{}c@{}}Approximate\\ Sensor Cost\end{tabular} \\ \midrule
Eddy Current Analysis\cite{eddy}                                                                              & Inner Race                                                                                 & Vibration                                                                 & Eddy Current Sensor                                                                           & Medium                                                                               & FFT                                                                                            & 5                                                                        & \$200                                                             \\ \\
Flux Analysis\cite{flux}                                                                                      & Bearing Defect                                                                             & Stray Flux                                                                & Magnetic Flux Sensor                                                                          & Medium                                                                               & FFT                                                                                            & 50                                                                       & \$100                                                             \\ \\
Sound Analysis\cite{sound1}                                                                                     & \begin{tabular}[c]{@{}c@{}}Broken Rotor Bar,\\ Bearing Defect and\\ Unbalance\end{tabular} & Sound                                                                     & Condensor Microphone                                                                          & Low                                                                                  & MUSIC                                                                                          & 8                                                                        & \$100                                                             \\ \\
Phase Locked Loop (PLL)\cite{radar}                                                                            & Rotor and Bearing Defect                                                                   & Vibration                                                                 & Ultra Wideband Radar                                                                          & High                                                                                 & PLL                                                                                            & 2.5                                                                      & \$50                                                              \\ \\
Optical Doppler Shift\cite{optical1}                                                                               & Test Speaker Vibrations                                                                    & Vibration                                                                 & Laser Module                                                                                  & High                                                                                 & HHT                                                                                            & NA                                                                       & \$1000                                                            \\  \\
\begin{tabular}[c]{@{}l@{}}Multi Sensor Wireless \\ System\cite{vca}\end{tabular}                            & \begin{tabular}[c]{@{}c@{}}Bearing Defect and Air-\\ Gap Eccentricity\end{tabular}         & \begin{tabular}[c]{@{}c@{}}Acoustic, Vibration\\ and Current\end{tabular} & \begin{tabular}[c]{@{}c@{}}Hall Effect Sensor,\\ Accelerometer (2-axial)\\ and Microphone\end{tabular} & Medium                                                                               & FFT, HHT                                                                                       & 4                                                                        & \$500                                                             \\ \\
\begin{tabular}[c]{@{}l@{}}Motor Current Signature\\ Analysis\cite{mcsa}\end{tabular}                         & Bearing Defect                                                                             & Current                                                                   & Current Sensor                                                                                & Medium                                                                               & Wavelet Decomposition                                                                          & NA                                                                       & \$100                                                             \\ \\
Speed Analysis\cite{rotor}                                                                                     & Bearing Defect                                                                             & Rotor Speed                                                               & \begin{tabular}[c]{@{}c@{}}E60H NPN Type Rotary\\ Encoder\end{tabular}                        & Medium                                                                               & \begin{tabular}[c]{@{}c@{}}Absolute Value Based\\ Principal Component \\ Analysis\end{tabular} & NA                                                                       & \$500                                                             \\ \\
\begin{tabular}[c]{@{}l@{}}Antenna's Near Reactive Field\\ Approach (Proposed Method)\end{tabular} & \begin{tabular}[c]{@{}c@{}}Bearing Defect and \\ Unbalance\end{tabular}                    & Vibration                                                                 & Omnidirectional Antenna                                                                       & Low                                                                                  & FFT, DCNN                                                                                      & 3                                                                        & \$2                                                               \\ \bottomrule
\end{tabular}
}
\label{comparison}
\end{table*}

\begin{table*}[!h]
\centering
\caption{Complexity Analysis}
\label{complex}
\renewcommand{\arraystretch}{2}
\resizebox{0.9\textwidth}{!}{%
\begin{tabular}{|l|c|c|c|c|c|c|c|c|c|c|c|c|c|}
\hline
S.No              & Depth             & Input Size        & \multicolumn{9}{c|}{Accuracy (\%)}                                                                  & \multicolumn{2}{c|}{Complexity}                  \\ \hline
\multirow{2}{*}{} & \multirow{2}{*}{} & \multirow{2}{*}{} & \multicolumn{3}{c|}{S11 Magnitude} & \multicolumn{3}{c|}{S11 Phase} & \multicolumn{3}{c|}{Combined} & FLOPs                        & No. of Parameters \\ \cline{4-14} 
                  &                   &                   & At 0 cm   & At 5 cm   & At 10 cm   & At 0 cm  & At 5 cm  & At 10 cm & At 0 cm  & At 5 cm & At 10 cm &                              &                   \\ \hline
1                 & 3                 & $150\times150\times3$         & 95.8      & 91.2      & 91         & 91.7     & 90.5     & 86.9     & 97.9     & 96.2    & 92.6     & 2.99 x $10^{8}$ & 112,320           \\ \hline
2                 & 3                 & $100\times100\times3$         & 96.0      & 93.6      & 90.2       & 92.1     & 90.9     & 87.4     & 98.2     & 96.7    & 92.2     & 1.56 x $10^{8}$ & 112,320           \\ \hline
3                 & 3                 & $50\times50\times3$           & 81.6      & 78.4      & 75.2       & 79.5     & 76.8     & 74.3     & 86.7     & 83.1    & 79.8     & 3.98 x $10^{7}$ & 112,320           \\ \hline
\end{tabular}
}
\end{table*}

Table \ref{comparison} shows the comparison of existing methods of fault detection with the proposed one based on various aspects. It is seen that most of the methods employing sensors like flux sensor, eddy current sensor, rotary encoder, etc. are expensive compared to the antenna used in our work. Most expensive method includes sensors such as laser module, accelerometer (2,3-axial) and spectrometer. Most of the sensors have their own electronic circuitry which makes them vulnerable to high working temperatures and circuit failures. However, the antenna requires only radio frequency sources and offers robust performance while it is placed near the faulty machines irrespective of the working environment. The antenna’s reactive near field can only be perturbed by being inside the range of reactive near field and thus it is more robust and less susceptible to environmental factors than other methods. Moreover, the proposed method is based only the vibration signal unlike other methods where they use combination of different signals (current, speed, acoustic and vibration signal). 
The proposed method has advantages such as low sensor complexity, is cost-effective, temperature resistant, low power consumption and low signal duration (3 sec) is needed for the analysis. The method demonstrates its ability to effectively classify the faults using DCNN. Furthermore, due to the effect of noise and other environmental factors, as well as the lack of phase data in most of the methods, applying a similar DCNN model to the spectrograms produced from other methods will not be much noticeable, resulting in limited classification performance.

\section{Complexity Analysis}
The model complexity is also evaluated by analyzing its time and space complexity \cite{complex}. The time complexity indicates the number of operations that needs to be performed to complete a task, while the number of parameters required by a model is referred to as space complexity. They are expressed as,
\begin{flalign}
&Time \;  complexity \sim O\left (   \sum_{i=1}^{N} n_{i-1} \cdot  m_{i}^{2} \cdot k_{i}^{2} \cdot n_{i} \right )& 
\end{flalign}
\begin{flalign}
&Space \;  complexity \sim O\left (   \sum_{i=1}^{N} n_{i-1} \cdot k_{i}^{2} \cdot n_{i} \right )&
\end{flalign}

where $N$ is the number of convolution layers, $n_{i-1}$ and $n_{i}$ is the number of channels at $i-1$ and $i^{th}$ layer respectively. The size of feature map is denoted by $m^{2}_{i}$ and the size of the kernel is denoted by $k^{2}_{i}$, at $i^{th}$ layer. The above equation does not account for the time cost of fully connected layers and pooling layers. As these layers often occupy 5-10\% of the total computational time, only the complexity of the convolutional layers is considered.                                                                                       

Table \ref{complex} shows the model accuracy and complexity for three different input sizes. The accuracy is computed at 5.8 GHz. It is found that $112,320$ parameters are required by the model. For an input size of $100\times100\times3$, the model has achieved the highest accuracy with $1.56 \times 10^{8}$ floating-point operations per second (FLOPs). For input size of $150\times150\times3$, the model reported similar accuracy but with $2.99 \times 10^{8}$ FLOPs which is higher compared to others. For an input size of $50\times50\times3$, the model reported the lowest FLOPs with $3.98 \times 10^{7}$ at the cost of poor accuracy. The complexity and accuracy achieved is best when the input size is $100\times100\times3$.

\section{Conclusion}
In this paper, a single antenna-based noncontact and cost-effective method is proposed and implemented successfully to classify different faults such as inner race fault, outer race fault and imbalance in induction motor using the antenna's near reactive field. The vibration due to the various fault disrupts the near reactive field of the antenna and as a result, causes a variation in the measurement of the antenna's reflection coefficient. We have validated the antenna's performance in picking up the vibration signals due to specific faults based on FFT analysis and average power analysis.

A deep convolutional neural network is designed and applied to the spectrograms of the reflection coefficient to classify different faults. It is found that the 5.8 GHz frequency provides the highest classification accuracy among the three frequencies tested (433 MHz, 2.4 GHz, and 5.8 GHz) using both magnitude and phase of the reflection coefficient. The effect of antenna distance from the vibration source is also studied by considering three different antenna positions, and it is found that the accuracy is highest at 5.8 GHz among the three frequencies. It is worth noting that as the operating frequency increases, so does the range of the near reactive field which may get disrupted by other nearby machines. Thus, it is best to select the operating frequency based on the range of the near reactive field. Furthermore, the impact of the S11 signal's duration on classification accuracy is also investigated and it has been found that for the best classification accuracy, input data must be at least 3 seconds long. The complexity of the model is also analyzed. Our result shows the effectiveness of the antenna as a sensor in recognizing different operating faults which can be used for early fault diagnosis in industries.

Future research could include measurements with a variety of antennas and diverse antenna placements to identify vibration related faults. More operational settings could be included to better assess the applicability of the suggested technique. Finally, more advanced signal processing methods can be used to retrieve signal data, potentially to improve the classification accuracy and computational complexity for real-time applications.

\bibliographystyle{IEEEtran}
\bibliography{reference}


\end{document}